\documentclass[conference,letterpaper]{IEEEtran}

\usepackage{xspace}
\usepackage{amsmath,amsfonts}
\usepackage{algorithmic}
\usepackage{graphicx}
\usepackage{textcomp}

\def\BibTeX{{\rm B\kern-.05em{\sc i\kern-.025em b}\kern-.08em
    T\kern-.1667em\lower.7ex\hbox{E}\kern-.125emX}}

\usepackage{tikz}
\usepackage{amsmath}

\usepackage{eucal}
\usepackage{filecontents}
\usepackage{epsfig,endnotes}
\usepackage{xspace}
\usepackage{amsmath}

\usepackage{pifont}
\usepackage{todonotes}
\usepackage[normalem]{ulem}
\usepackage{balance}
\usepackage{gensymb}
\usepackage{multirow}
\usepackage{subcaption}
\usepackage{caption}
\usepackage{mathtools}
\usepackage{bbm}
\usepackage{balance}
\usepackage[keeplastbox]{flushend}
\usepackage{booktabs}
\usepackage{makecell}

\newcommand{\cmark}{\ding{51}}%

\newcommand{\eg}{e.g., \xspace}
\newcommand{\ie}{i.e., \xspace}

\usepackage[export]{adjustbox}

\usepackage{cellspace}
\newcommand\blfootnote[1]{%
  \begingroup
  \renewcommand\thefootnote{}\footnote{#1}%
  \addtocounter{footnote}{-1}%
  \endgroup
}

\usepackage[export]{adjustbox}

\begin{document}
	
\title{Understanding Worldwide Private Information Collection on Android}

\author{\IEEEauthorblockN{Yun Shen}
\IEEEauthorblockA{NortonLifeLock Research Group\\
yun.shen@nortonlifelock.com}
\and
\IEEEauthorblockN{Pierre-Antoine Vervier}
\IEEEauthorblockA{NortonLifeLock Research Group\\
pierre-antoine.vervier@nortonlifelock.com}
\and
\IEEEauthorblockN{Gianluca Stringhini}
\IEEEauthorblockA{Boston University\\
gian@bu.edu}}

\maketitle

\begin{abstract}
\blfootnote{This paper appeared in the 2021 ISOC Networks and Distributed Systems Security Symposium}Mobile phones enable the collection of a wealth of private information, from unique identifiers (e.g., email addresses), to a user's location, to their text messages.
This information can be harvested by apps and sent to third parties, which can use it for a variety of purposes.
In this paper we perform the largest study of private information collection (PIC) on Android to date.
Leveraging an anonymized dataset collected from the customers of a popular mobile security product, we analyze the flows of sensitive information generated by 2.1M unique apps installed by 17.3M users over a period of 21 months between 2018 and 2019.
We find that 87.2\% of all devices send private information to at least five different domains, and that actors active in different regions (e.g., Asia compared to Europe) are interested in collecting different types of information.
United States (62\% of the total) and China (7\% of total flows) are the largest two countries that collect most of the private information.
Our findings raise issues regarding data regulation, and would encourage policymakers to further regulate how private information is used by and shared among the companies and how accountability can be truly guaranteed.
\end{abstract}

\section{Introduction}
\label{sec:introduction}

Data has become the commodity that sustains much of the Web.
In recent years, the research community has raised awareness on the threats linked to sensitive user data collection by third parties.
For example, specialized companies collect information from Web users to uniquely identify them across websites, potentially to provide them with more tailored advertisements~\cite{acar2013fpdetective,iordanou2018tracing,lerner2016internet,mayer2012third,starov2017extended}.
In some cases, rogue browser extensions collect information that is supposed to remain private, such as a user's browsing history~\cite{chen2018mystique,weissbacher2017ex}.
As mobile devices become more central in the computing experience of users, the threats linked to private information collection increase.
Mobile devices can provide a wealth of sensitive information~\cite{leung2016should,papadopoulos2017long} that goes beyond identifiers to uniquely fingerprint users~\cite{razaghpanah2018apps}, including location information~\cite{fawaz2014location,ma2013detecting,vanrykel2016leaky}, call logs, text messages, and even information on which applications are installed a device~\cite{zhou2011taming}.
This information can be used by third parties to deliver targeted advertisements~\cite{nath2015madscope} as well as for nefarious reasons, from stalking a victim by monitoring her location~\cite{chatterjee2018spyware} to defeating two factor authentication by stealing text messages~\cite{kaspersky2factor}.

There exist insightful research efforts~\cite{binns2018third,continella2017obfuscation,demetriou2016free,gibler2012androidleaks,iordanou2018tracing,joe2015sponsoring,liu2019privacy,razaghpanah2018apps,ren2016recon,seneviratne2015measurement,stevens2012investigating,vallina2012breaking} to understand the impact and the threats posed by information collection on mobile devices.
It remains however very challenging to obtain a comprehensive view of the information collected by mobile apps, given the wealth of potential information collected, the software diversity of mobile platforms, and the geographic diversity of mobile users and of the actors that they interact with.
To shed light on the problem, previous research resorted to running apps in a sandbox environment~\cite{continella2017obfuscation} or analyzing network traffic~\cite{binns2018third,ren2016recon,seneviratne2015measurement} to monitor the information that they leaked. 
While this approach can be useful to identify trackers, it has two limitations: first, by running apps from a single vantage point it is challenging to replicate the geographic diversity of real users; second, apps could detect sandboxed environments and act in a different way than they would on real devices (for example by not leaking any sensitive information), and this could bias the results~\cite{miramirkhani2017spotless,rossow2012prudent,vidas2014evading}.
Alternatively, previous work collected network data from an ISP, looking for information leaks~\cite{iordanou2018tracing,joe2015sponsoring,vallina2012breaking}. 
While this approach solves the sandbox detection problem, it still has a geographic bias, since different users around the world might be using different apps and might be subject to different types of sensitive data collection.
As a third approach, researchers recruited participants to install an app on their mobile phones; the app would then monitor the device for information leaks~\cite{razaghpanah2018apps}.
This approach solves the issues mentioned above and offers insightful findings, but it remains a challenging task to attract a population of users that is large and diverse enough to represent worldwide trends at scale.
Additionally, previous research~\cite{razaghpanah2018apps} either mainly looked at information collections that can be used to identify a device (\eg IMEI numbers or SIM card information) or considered limited types of sensitive information that can be collected by third parties~\cite{leung2016should,pan2018panoptispy,ren2016recon}, such as birthday, username/passwords, contacts, media files, etc.

In this paper, we provide the most comprehensive view of private data collection by Android apps to date.
To achieve this, we tap into the analysis infrastructure of a popular mobile security product.
The company behind this product runs Android apps in its backend infrastructure and identifies dangerous information flows by performing static and dynamic analysis.
It then builds signatures of method calls that are indicative of privacy invasive activity and pushes them to the mobile devices that installed the security product, which use them to identify privacy invasive and malicious apps that have been installed.
This infrastructure allows us to monitor the information collected by apps for a population of 17.3M devices daily for 21 months between 2018 and 2019.
This is three orders of magnitude more devices than what previous work analyzed~\cite{razaghpanah2018apps}.
Compared to previous work, we go beyond tracking, contact, and credential information, and trace 22 categories of private information, 13 of which were not considered by previous work~\cite{binns2018third,pan2018panoptispy,razaghpanah2018apps,ren2016recon} (see Section~\ref{sec:datasets}). This allows us to paint an unprecedented picture of the state of sensitive information collection on Android in the wild, identifying the big players in this space (both legitimate companies and malicious actors), together with geographic trends.

\begin{table}
\centering
\caption{Summary of datasets used.}
\resizebox{0.8\linewidth}{!} {
\begin{tabular}{llr}    \toprule
\textbf{Dataset} & \textbf{Data} & \textbf{Count}  \\
\midrule
Mobile app activity log    & Total records  & 6B \\
(01/2018 - 09/2019) & Days           & 634 \\
 & Countries and regions & 201 \\
 & Devices               & 17.3M \\
 & Distinct app names  & 2.13M    \\ 
 & Distinct app SHA2s  & 6.5M  \\
 & Distinct PIC FQDNs & 76,451 \\
 & Distinct PIC domains & 40,851 \\

Mobile app reputation log & Low reputation SHA2s & 3.4M \\
\midrule
VT                 & Total reports           & 6.5M \\
 & PHA SHA2s (detections $\geq 6$)      & 3.5M \\
 & Benign SHA2s (no detection)  & 2.3M \\
 & Not found SHA2s & 401K \\
\midrule
Domain to owner org. & Domains  & 10,736 \\
(01/2018 - 09/2019) & Organizations & 9,593 \\
\midrule
Blacklists & Domains/IPs & 7,670 \\
(01/2018 - 09/2019) & & \\
\midrule
Geolocation & Domains/IPs & 40,851 \\
(01/2018 - 09/2019) & & \\
\bottomrule
\end{tabular}
} 
\label{tab:datasets}
\end{table}

\noindent Among others, this paper makes the following findings:
\begin{itemize}
 \item Private information collection is widespread on Android, with 87.2\% of all devices in our dataset sending information to at least five distinct domains.
   While most PIC domains collect identifiers to track a user or a device (\eg device information or email addresses), an alarmingly high number of domains collect other types of private information such as a device's location or a user's contacts.
 \item Looking at the destinations where private information is sent to, we find that most information flows terminate in the United States.
   We do however find that China, trailing the US at the second place, collects 7\% of all data flows.
   This is three times higher than what was reported in previous work~\cite{razaghpanah2018apps}.
   We also find that there was no significant difference in the number of information flows leaving the European Union after the implementation of GDPR.
   These findings highlight the challenges involved in implementing data protection regulations.
 \item We find that potentially harmful applications (PHAs)~\cite{googlereport} are more aggressive in collecting private information than benign apps, especially when it comes to information related to the apps installed and running on a device.
   We also find that a small number of devices (4k) had apps installed that steal the user's text messages, potentially enabling the circumvention of two factor authentication.
\end{itemize}

Our findings highlight a number of challenges faced by the research community when studying private information collection on Android. 
We show that looking at device penetration is critical to observe the distribution of information collection actors in the wild, and looking at application penetration only can provide a biased view.
We also highlight how looking at users located in different regions is important to get a comprehensive view, since actors operating in different countries are interested different types of information.

\section{Datasets}
\label{sec:datasets}

\begin{figure}
\centering
\includegraphics[width=0.9\linewidth]{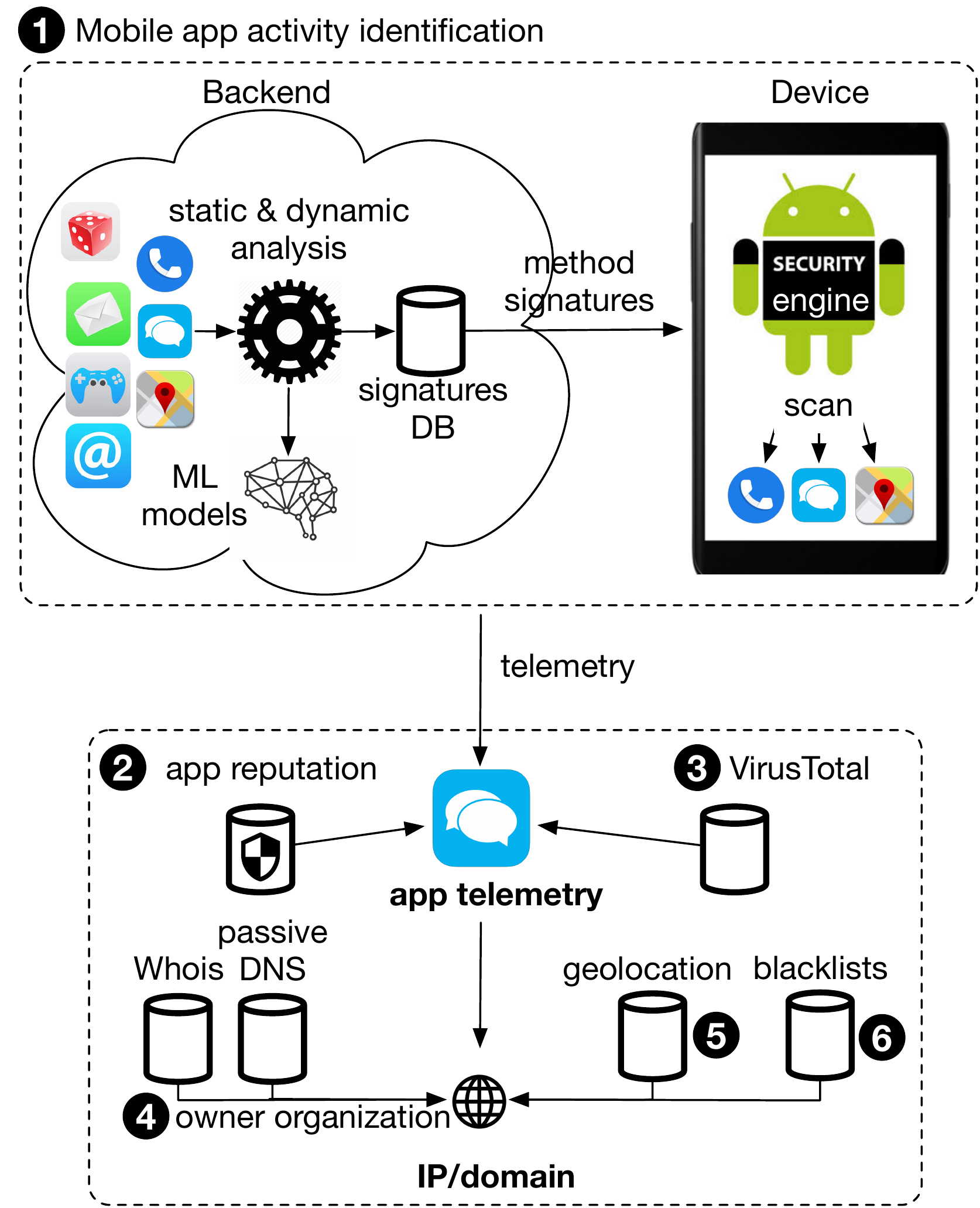}
\caption{Workflow of our measurement study.}
\label{fig:workflow}
\end{figure}

\begin{table*}[t]
\centering
\caption{The 22 types of private information monitored by the security app. 
}
\resizebox{0.9\linewidth}{!} {
\begin{tabular}{llll}
\toprule
\textbf{Group} & \textbf{Category} & \textbf{Description} & \textbf{Previously studied or novel}  \\
\midrule
Tracking & Phone number & Phone number & \cite{leung2016should,razaghpanah2018apps} \\
 & Device info &  IMEI, OS/kernel version, phone producer, phone model & \cite{leung2016should,razaghpanah2018apps,ren2016recon}\\
 & SIM card info & Information about SIM serial number, IMSI, voicemail number  &   \cite{razaghpanah2018apps} \\
 & Location info & GPS or cell tower coordinates    & \cite{leung2016should,ren2016recon} \\
 & Operator info & Information about the network operator   & \cmark \\
 & Setting info  & Information about the device configurations  &  \cmark \\
\midrule
Activity and social profiling & Account info  & \begin{tabular}[c]{@{}l@{}}Details about the configured accounts can be exported \\ (including user names of entries under Settings/Accounts)\end{tabular}   & \cite{leung2016should,ren2016recon} (partially)\\
 & Email info & Details about the email address such as Gmail address can be exported  &  \cite{leung2016should} \\
 & Contact info  & Contact list  can be exported  & \cite{ren2016recon}\\
 & Social network account  & Details about the social network accounts such as Facebook account can be exported  & \cmark \\
 & Voice mail account  & Details about the voice mail accounts can be exported   &  \cmark \\
 & Call log & Call log can be exported  &  \cmark \\
 & SMS info & App can send the content or sender/recipient details from SMS/MMS messages  &  \cmark \\
 & Calendar info& Calendar can be exported  & \cmark \\
\midrule
Usage preference & Installed app info & \begin{tabular}[c]{@{}l@{}}Details about apps installed on the phone are/can be exported\\ (full or partial list of installed package names, or app titles)\end{tabular}  &  \cmark \\
 & Running app info & Details about apps running at a certain time are/can be exported  &  \cmark \\
 & Browser history info  & Browser history can be exported  &  \cmark \\
 & Browser bookmark info  & Browser bookmarks can be exported   &  \cmark \\
\midrule
Audio/Video & Audio info & \begin{tabular}[c]{@{}l@{}}Recorded audio clips can be exported \\ (e.g., recorded by the app, or picked from saved)\end{tabular}   & \cite{pan2018panoptispy}  \\
 & Photo info  & Photo can be exported &  \cmark \\
 & Video info  & Video can be exported & \cite{pan2018panoptispy}   \\
 & Camera info & App can take pictures or picks them from gallery and exports them &   \cmark \\
\bottomrule
\end{tabular}
}
\label{tab:catsofpi}
\end{table*}

This section details the approach that we follow for data collection (summarized in Figure~\ref{fig:workflow}) and summarizes the datasets used in this study (see Table~\ref{tab:datasets}).

\noindent \textbf{Workflow.} 
The overall workflow of our measurement study is as follows (see Figure~\ref{fig:workflow}).
We use \emph{mobile app activity data (\ding{182})} to identify the private information collection activities from 2.13M apps (6.5M SHA2s) installed on 17.3M devices across 200 countries and regions.
We then augment this data by using \emph{Mobile app reputation data (\ding{183})} and \emph{VirusTotal (VT) reports (\ding{184})} to identify the potentially harmful apps (PHAs). 
Finally we use \emph{domain and IP Whois and passive DNS (\ding{185})} to extract domain ownership information (\eg parent company, business category, etc), \emph{IP and domain geolocation (\ding{186})} to identify the country where apps send data, and \emph{IP and domain blacklists (\ding{187})} to identify domains associated with malicious activity.
We provide details of each step in the rest of this section.

\noindent \textbf{Telemetry data collection.}
In this paper, we use mobile telemetry data collected by the security company's mobile security product, which has been installed on millions of mobile devices. 
This company has a dedicated infrastructure to collect apks (one app may have multiple apks) from popular Android markets and various intelligence sources. 
These apks are then analyzed by a sophisticated infrastructure with both static and dynamic analysis pipelines. 
For instance, the static analysis pipeline can identify if an apk directly invokes any suspicious and sensitive API (including reflection~\cite{aafer2013droidapiminer}, dynamic code loading~\cite{poeplau2014execute}, native code~\cite{li2017static}, etc.), requests permissions not related to its advertised description~\cite{qu2014autocog}, as well as perform fine-grained permission analysis~\cite{sarma2012android,felt2011android}, flow and context sensitive taint analysis~\cite{arzt2014flowdroid}, etc.
Third-party libraries/SDKs used by the apps are also analyzed using the same procedure stated above. 
Following the static analysis, the backend can build an initial report on control-flow, data-flow, and permissions related to an apk.
In addition to static analysis, the security company also performs dynamic analysis by running an apk in a sandbox environment with various Android OS versions.
Through network and system instrumentation, the dynamic analysis pipeline runs an apk in different conditions (\eg UI-automation~\cite{hao2014puma}, input generation~\cite{wong2016intellidroid}, apk fuzzing~\cite{ye2013droidfuzzer}, etc.) with varied execution time to capture its activities under different contexts. 
For example, the dynamic analysis pipeline reports if advertisements appear outside of an app in unexpected places (\eg notification bar, shortcut, etc.) or exhibit unusual behaviors (\eg change the user's home page).
State-of-the-art commercial products are also employed by the security company to deal with challenges such as emulator/motion evasion, obfuscated code/libraries, etc. to assist the aforementioned analysis pipelines.
At the same time, several machine learning models are built using the features generated from the pipelines to enable the backend to detect sophisticated PHAs.
This way, the mobile security product can fingerprint activities with high accuracy and minimize false positives which may lead to an undesirable high customer churn rate.
Note that the infrastructure continuously inspects apks.
An apk that has been analyzed before may also be subject to regular reinspection.  
By combining the results of the static and dynamic analysis, the security company can rigorously fingerprint traces of app activity, including the types of private information that the app is collecting.
These traces are later used to develop signatures for the apps collecting private information, in the form of sequences of method signatures.

These signatures are then deployed in the security product on the mobile devices to identify installed apps that have been linked to private information collection.
If the user permits telemetry data collection, meta-information related to the app detections are sent to the telemetry data collection infrastructure and used to improve the app security features and its privacy leakage detection capability. 
The collected data is safeguarded by the global privacy policy of this security company. 
Devices are identified by a unique anonymized identifier, but it is not possible to link such an identifier back to the device. 
The mobile security app only collects detection metadata, and it cannot inspect network traffic data, hence the company does not collect any actual communication/user data, or other types of PII.
We provide a detailed discussion about ethics and data privacy at the end of this section. 

\section{Landscape of Private Information Collection (PIC) in Mobile Ecosystems}
\label{sec:pic_landscape}

In this section, we study the landscape of private information collection (PIC) in mobile apps. 
First, we look at the pervasiveness of PIC apps installed on the user base of the security vendor.
We then focus on the app presence rate (\ie the number of PIC domains in apps) to identify the global/regional top players.
We later focus on the device penetration rate (\ie the number of devices that a PIC domain collects information from) to uncover the most pervasive PIC domains globally as well as important regional players, understand what types of private information are collected by these PIC domains, and if we can observe behavioral differences in different regions regarding private information collection.

\subsection{Pervasiveness of PIC in Mobile Apps}
\label{sec:pic_pervasiness}

\begin{figure}[t]
     \centering
	 
    \begin{subfigure}[t]{0.22\textwidth}
        \includegraphics[width=\linewidth]{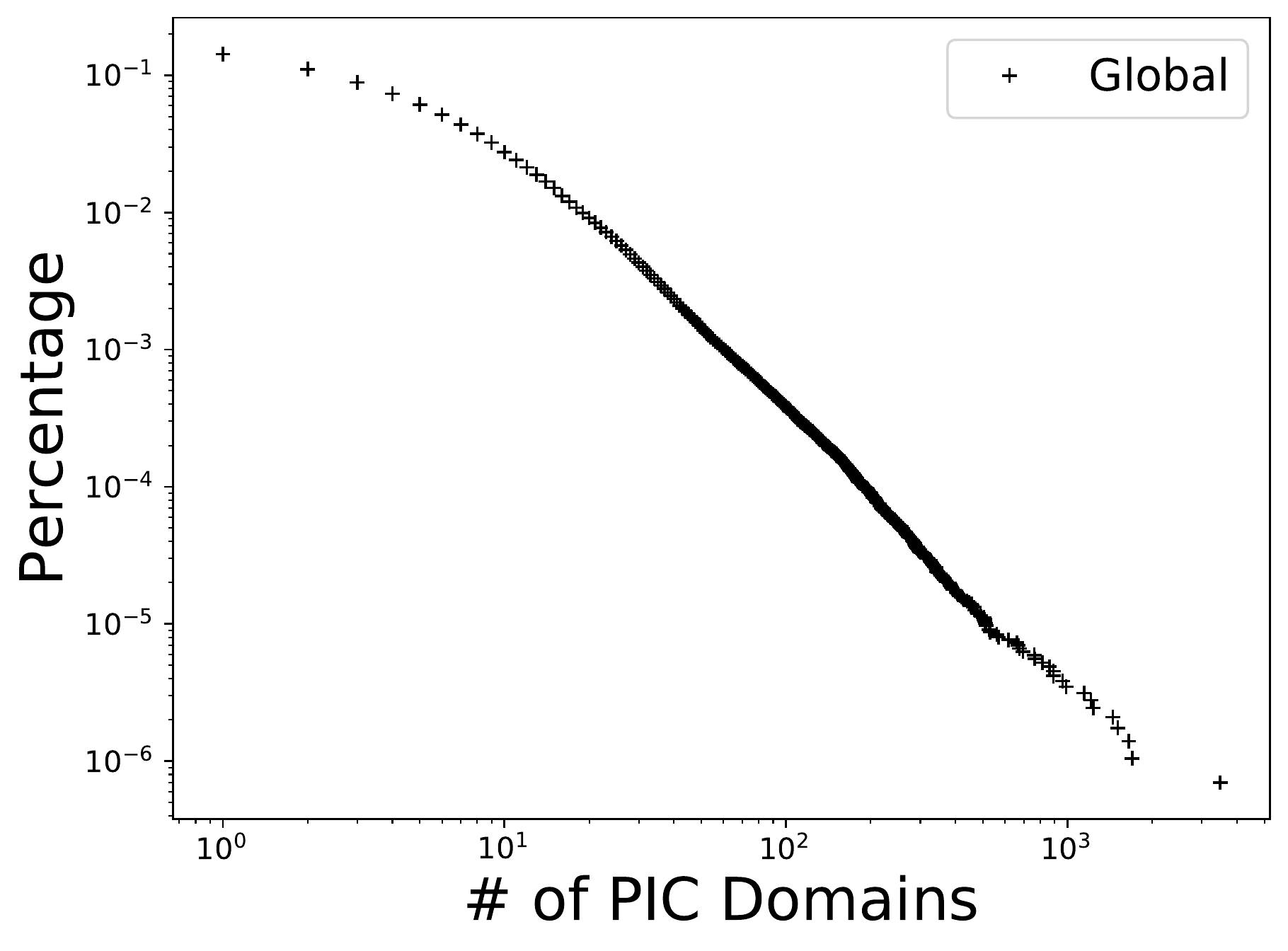}
        \caption{}
		\label{fig:app_pic_CCDF}
    \end{subfigure}
	\hfill
    \begin{subfigure}[t]{0.22\textwidth}
        \includegraphics[width=\linewidth]{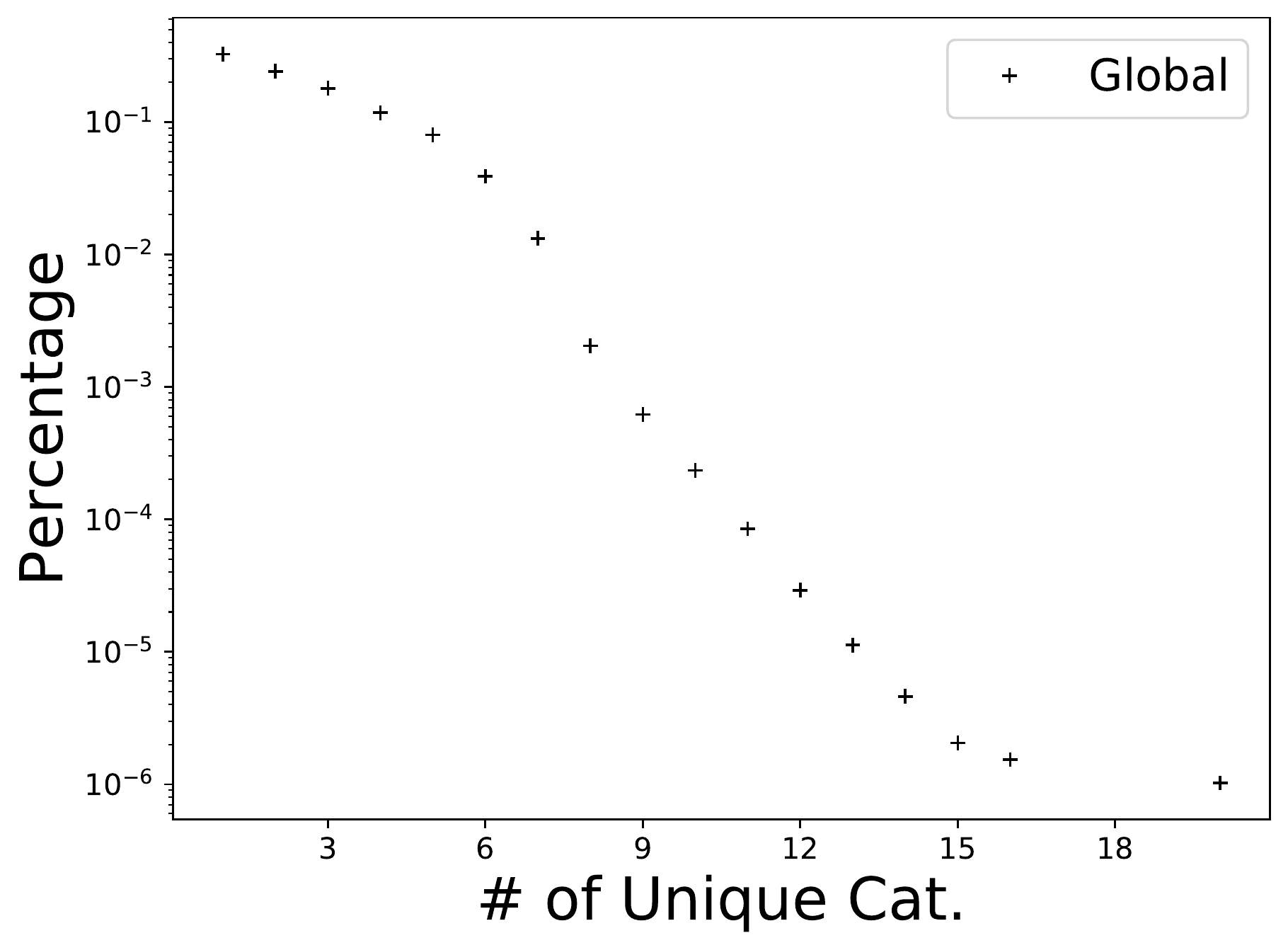}
        \caption{}
		\label{fig:app_cat_CCDF}
    \end{subfigure}
    \hfill
	\caption{\textbf{(log-scale)} Complimentary cumulative distribution (CCDF) of mobile apps in terms of unique PIC domains (a) and unique categories of private information collected (b).}
	\label{fig:pervasiveness}
\end{figure}

In this section, we demonstrate the pervasiveness of private information collection in mobile apps. 
The results are shown in Figure~\ref{fig:pervasiveness}. Apps send private information collected to 2 unique PIC domains on average. 
As we can observe in Figure~\ref{fig:app_pic_CCDF}, over 175K apps (approximately 8.2\% of total apps) send collected data to at least 5 unique PIC domains. 
These apps are installed on 15.1M devices in our dataset (87.2\% of all devices). At the same time, as we can observe in Figure~\ref{fig:app_cat_CCDF}, over 156K apps collect at least 5 unique categories of private information (see Table~\ref{tab:catsofpi}). 
This covers 13M devices (74.9\% of all devices). The overlapping 57.6k apps between the aforementioned two categories of apps cover 12.8M devices (73.8\% of all devices). In other words, 73.8\% of all devices in our dataset have at least one app collecting at least 5 unique categories of private information and sending them to at least 5 unique PIC domains. Our findings show that private information collection in mobile apps is universal and diversified at the same time.

\subsection{PIC Domains: App Presence Study}
\label{sec:pic_app_penetration}

\begin{figure}
\centering
\includegraphics[width=0.9\linewidth]{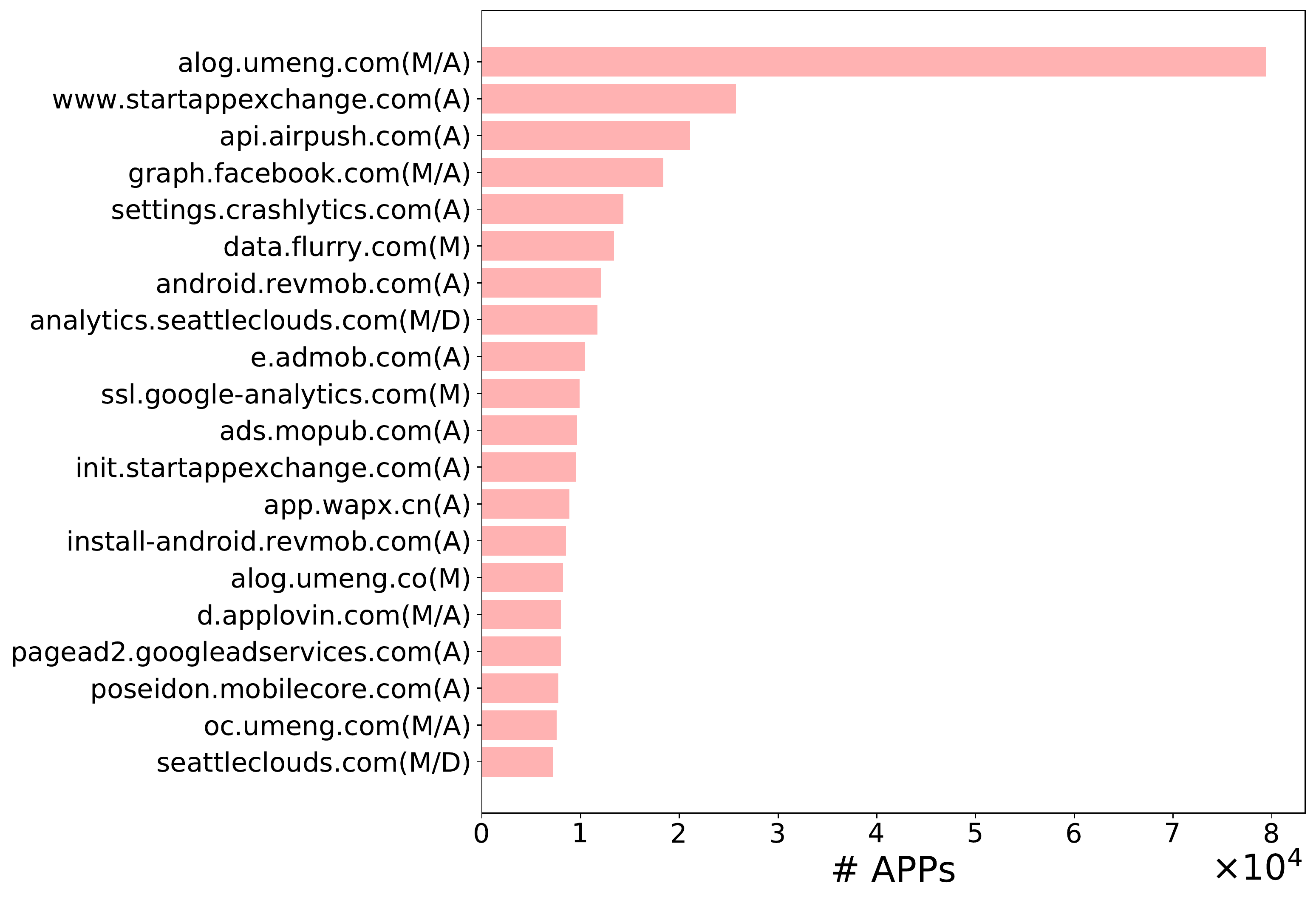}
\caption{Global top 20 PIC domains ranked by app presence. Domain's primary function- \textbf{M}: Metrics/Analytics, \textbf{A}: Advertising, and \textbf{D}: Development. }
\label{fig:dest_dist_all_by_app}
\end{figure}

\begin{figure*}
     \centering
\begin{adjustbox}{minipage=\linewidth,scale=0.83}
    \begin{subfigure}[t]{0.48\textwidth}
        \includegraphics[width=\textwidth,valign=b]{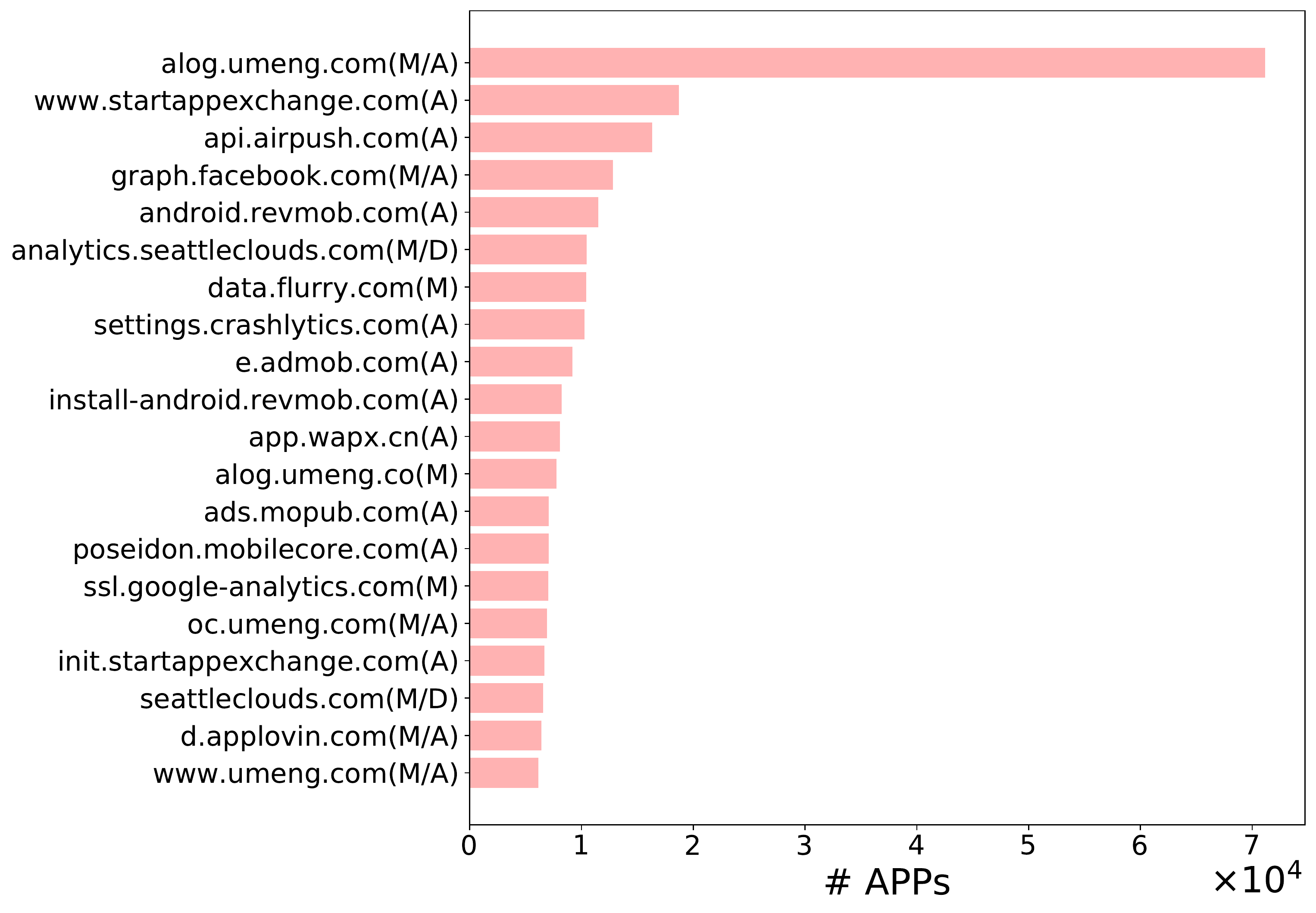}
        \caption{North America}
		\label{fig:dest_dist_northamerica_by_app}
    \end{subfigure}
	\hfill
    \begin{subfigure}[t]{0.48\textwidth}
        \includegraphics[width=\textwidth,valign=b]{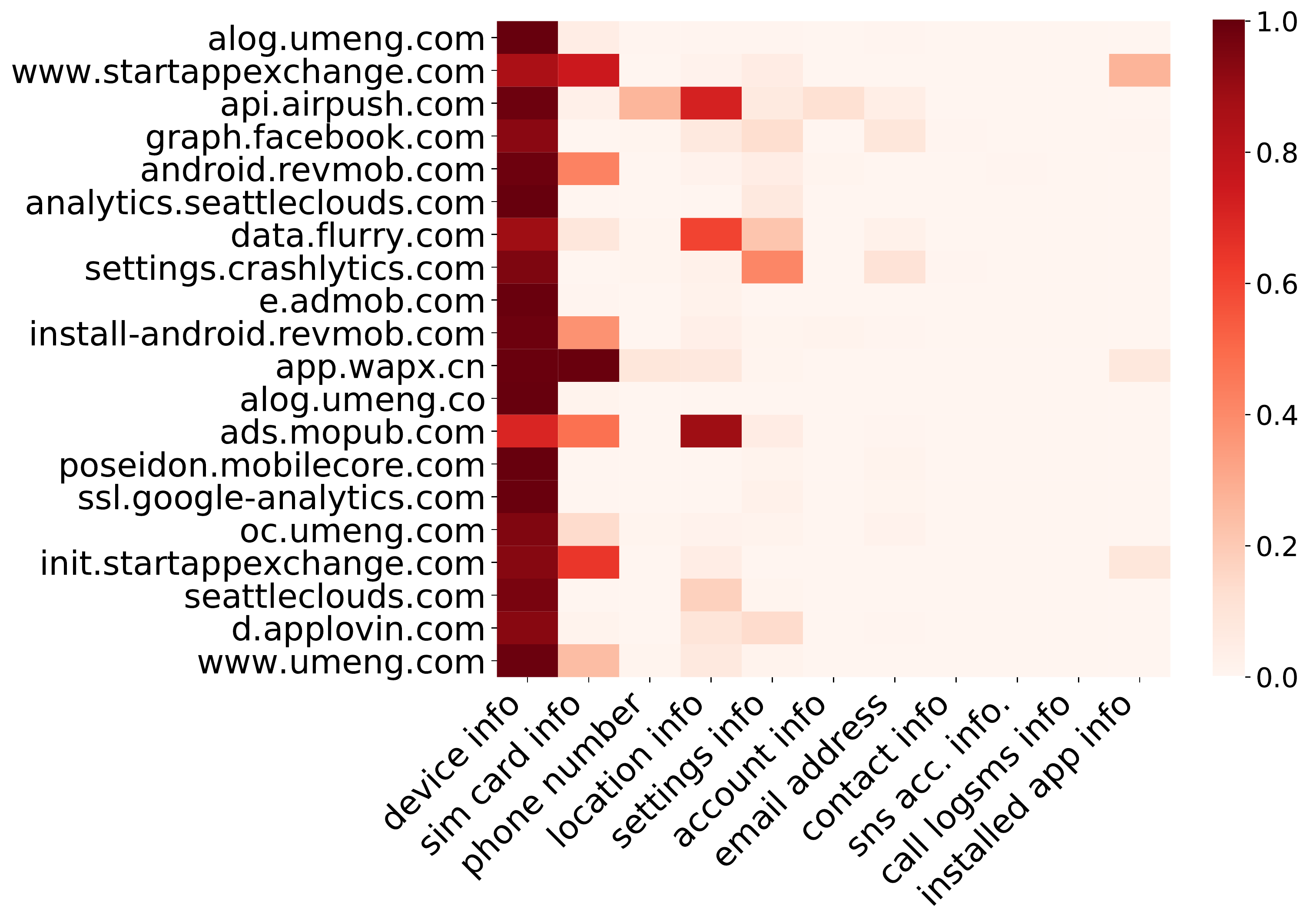}
        \caption{North America}
		\label{fig:heatmap_top_20_northamerica_by_device}
    \end{subfigure}
	
	\hfill
	
    \begin{subfigure}[t]{0.48\textwidth}
        \includegraphics[width=\textwidth,valign=b]{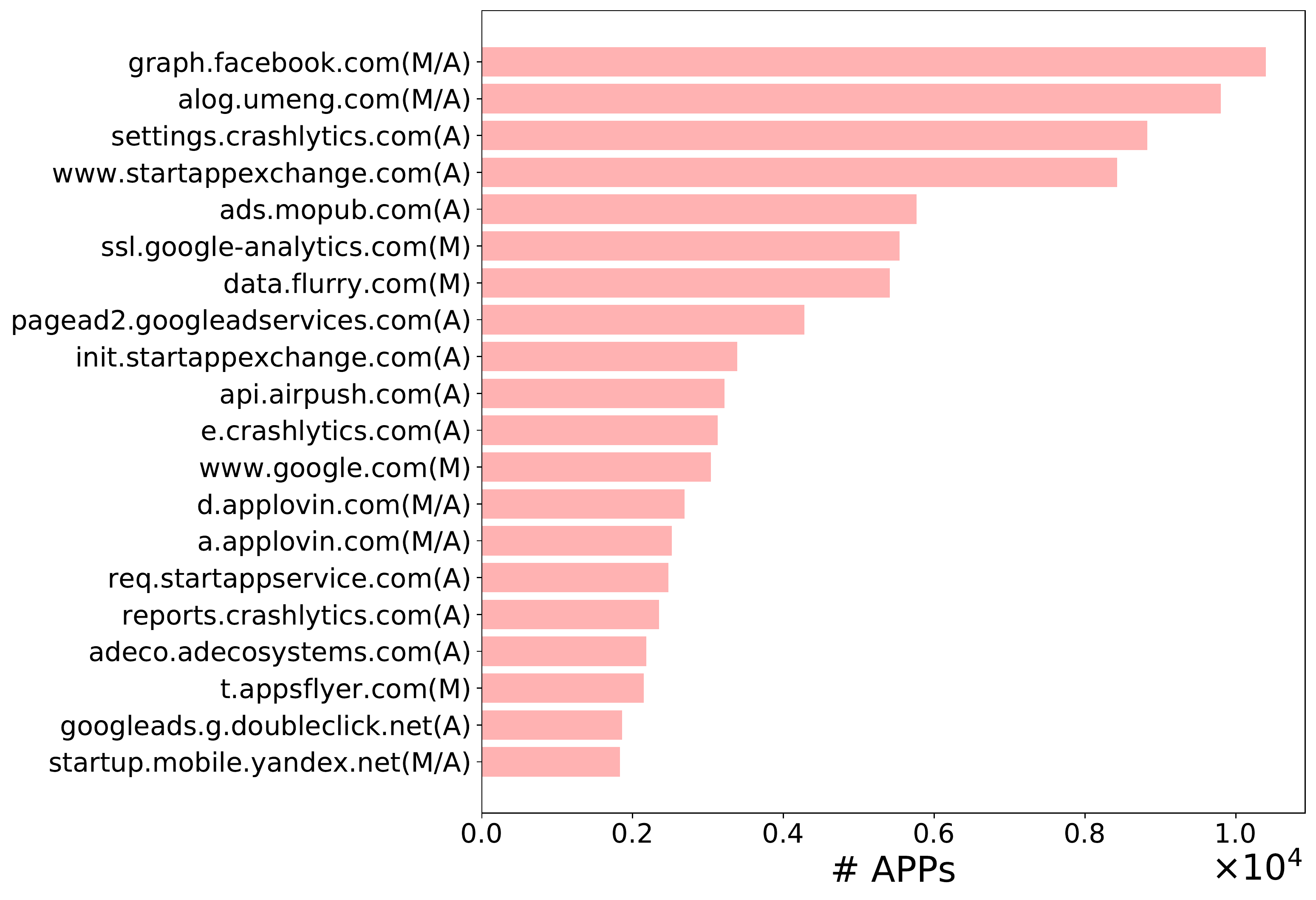}
        \caption{Europe}
		\label{fig:dest_dist_europe_by_app}
    \end{subfigure}
	\hfill
    \begin{subfigure}[t]{0.48\textwidth}
        \includegraphics[width=\textwidth,valign=b]{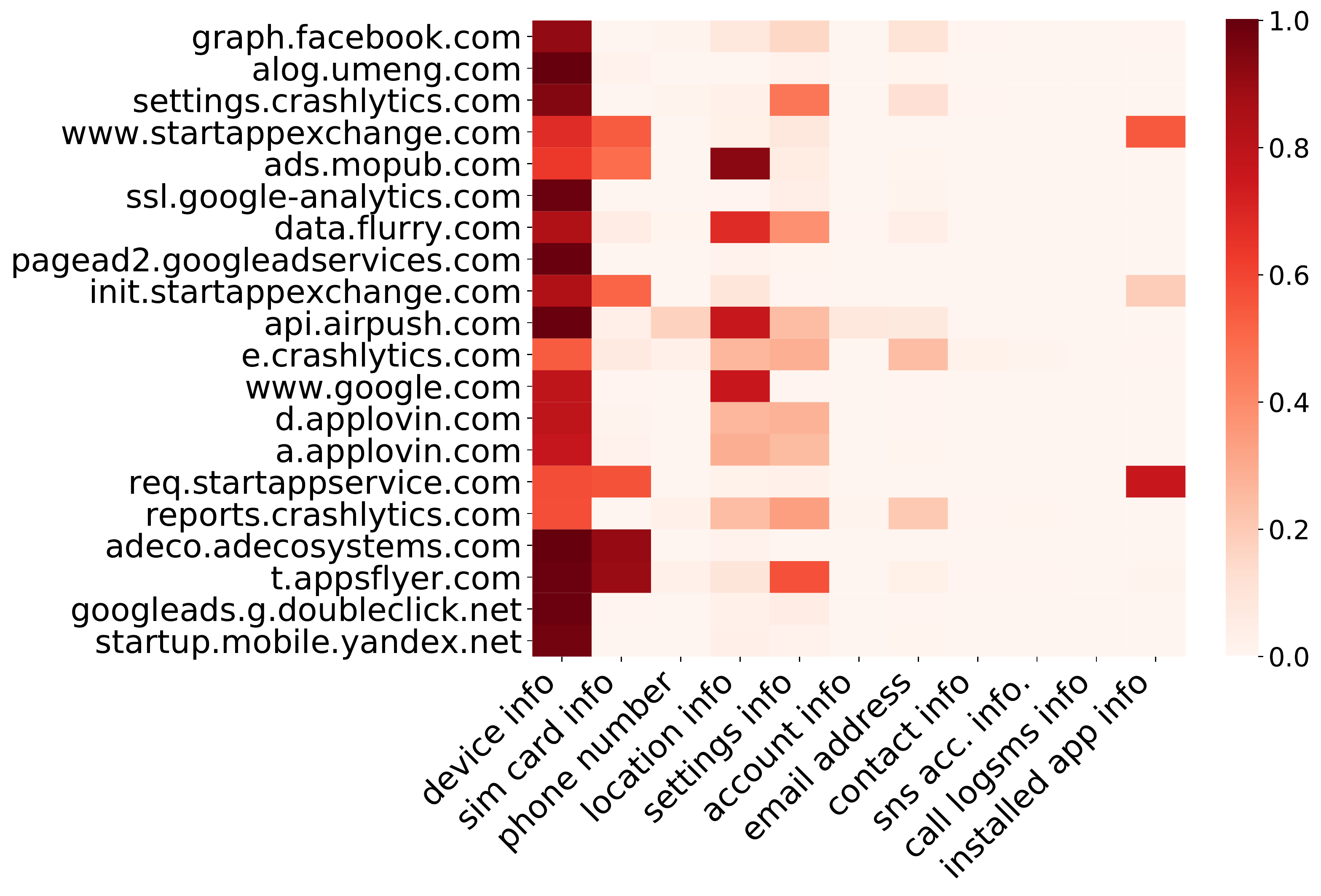}
        \caption{Europe}
		\label{fig:heatmap_top_20_europe_by_device}
    \end{subfigure}
	
	\hfill
    
	\begin{subfigure}[t]{0.48\textwidth}
        \includegraphics[width=\textwidth,valign=b]{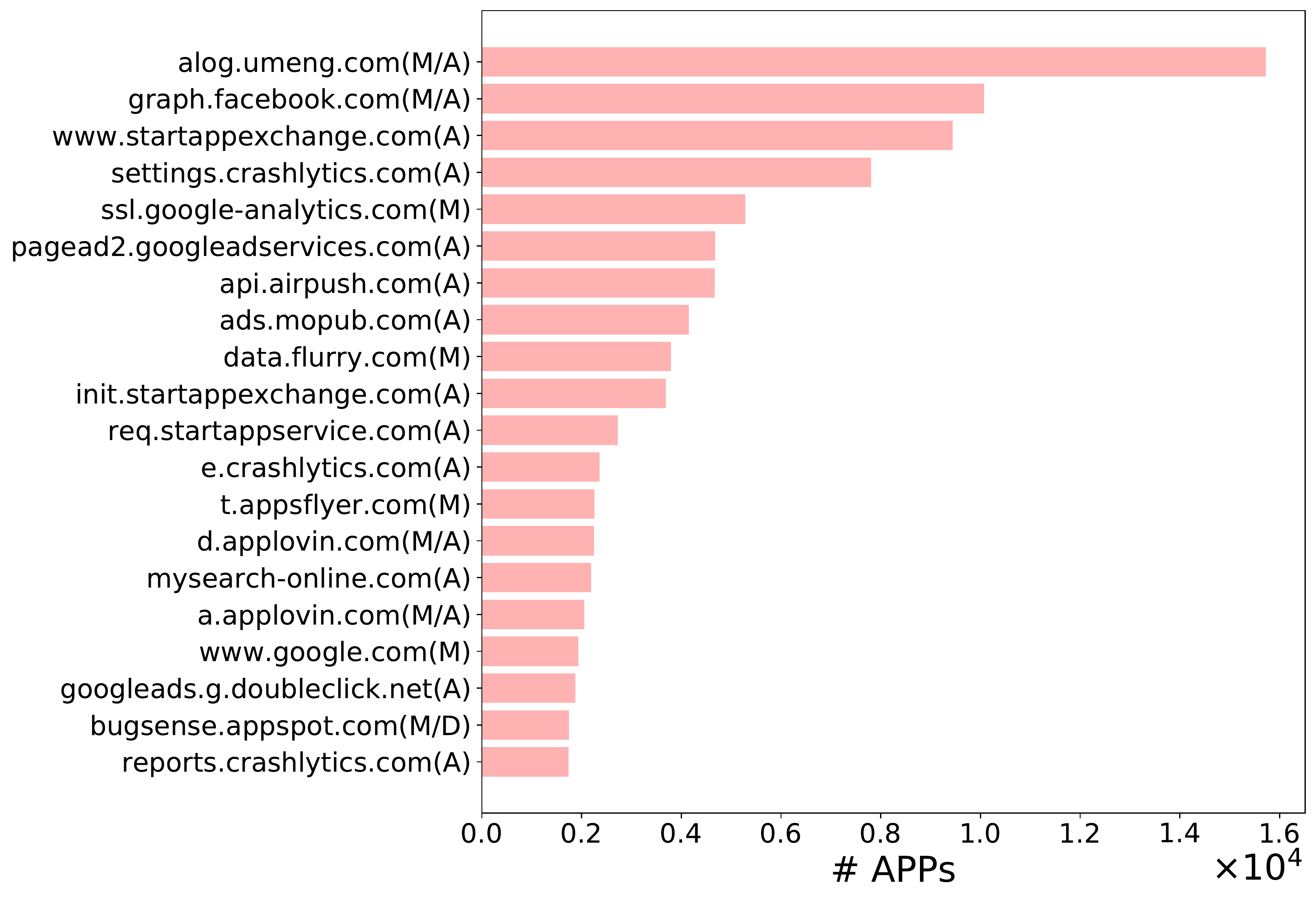}
        \caption{Asia}
		\label{fig:dest_dist_asia_by_app}
    \end{subfigure}
	\hfill
    \begin{subfigure}[t]{0.48\textwidth}
        \includegraphics[width=\textwidth,valign=b]{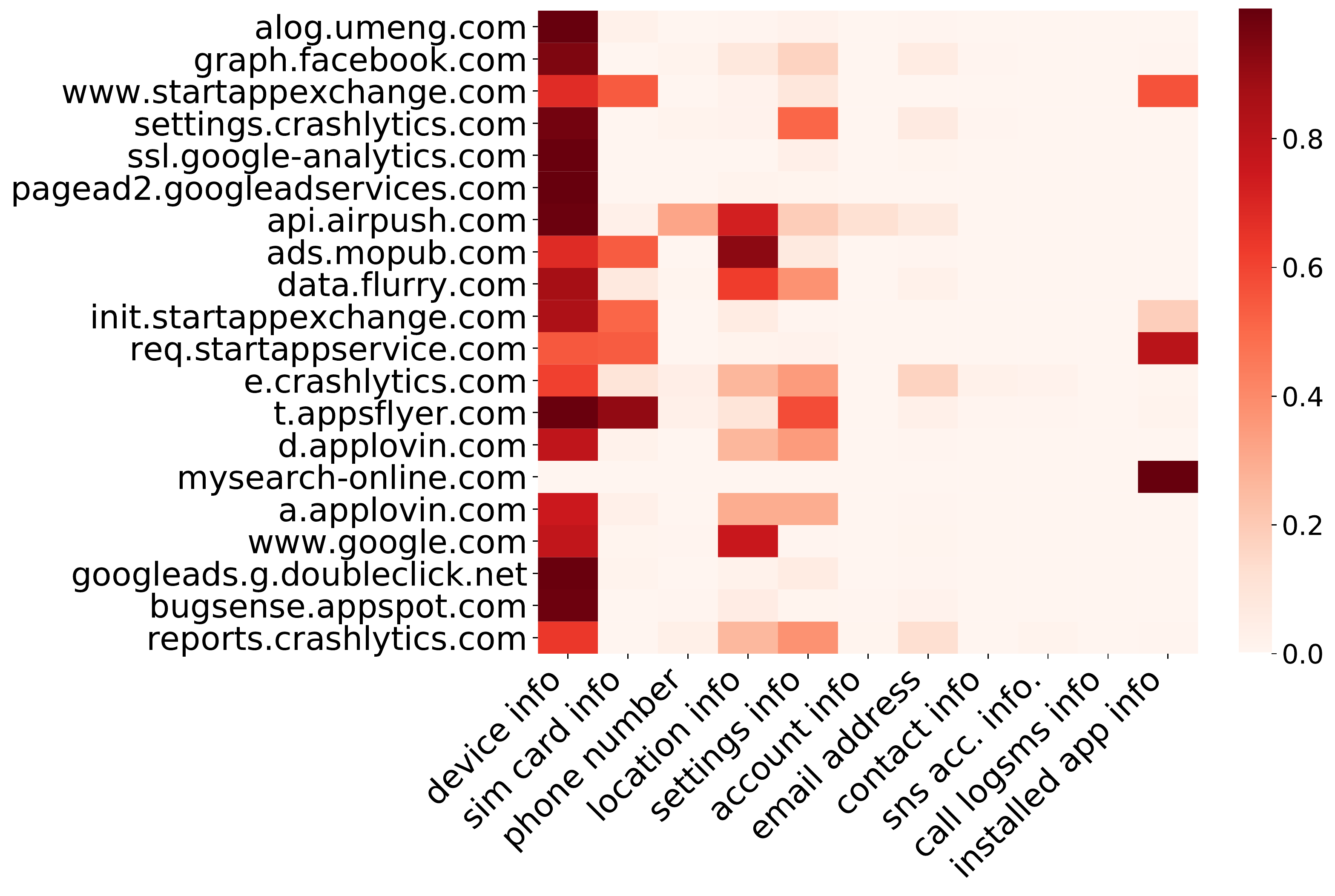}
        \caption{Asia}
		\label{fig:heatmap_top_20_asia_by_device}
    \end{subfigure}
	\hfill
    \caption{\textbf{(left column)} Regional top 20 PIC domains ranked by app presence. Domain's primary function - \textbf{M}: Metrics/Analytics, \textbf{A}: Advertising, and \textbf{D}: Development. \textbf{(right column)} Heatmap illustration of top 12 categories of private information collected by these PIC domains. Each row is normalized to [0, 1] by a PIC domain's total app presence. The darker the red implies that the more apps that a PIC domain collects information from.}
	\label{fig:dest_dist}
\end{adjustbox}
\end{figure*}

PIC organizations generally benefit from collecting data about more users. 
To reach this goal, one of the strategies adopted by these organizations is increasing their presence in mobile apps to reach out to more users. For example, an ad library will entice developers into including it in their apps. 
Figure~\ref{fig:dest_dist_all_by_app} shows the 20 PIC domains with the largest app presence globally (i.e., the domains that were contacted by the largest number of apps). 
Based on the information we collect from Crunchbase and the company websites, we attribute these PIC domains to three functions - Metrics/Analytics (M), Advertising (A), and Development (D). As we can see in Figure~\ref{fig:dest_dist_all_by_app}, the majority of these PIC domains (15 out of 20) offer advertising services. For example, in addition to the PIC domains owned by Google and Facebook, several known PIC domains operated by online advertisement companies (\eg \texttt{api.airpush.com}, \texttt{android.revmob.com}, \texttt{e.admob.com}, \texttt{ads.mopub.com}) have considerable global app presence, being contacted by 10K apps or more. Additionally, 8 out of the top 20 PIC domains offer metrics/analytics services. One noticeable finding from our study is \texttt{alog.umeng .com} (part of \texttt{Alibaba Group}). This domain has the largest global app presence and is contacted by 79,402 apps (3\% of total apps). 
This domain was not reported by previous measurement studies~\cite{razaghpanah2018apps}, and might be because our dataset contains three orders of magnitude more devices, distributed across the globe (recall that over 7M users are located in Asia).
Note that a high app penetration rate does not necessarily lead to a high device penetration rate, while the latter is directly proportional to the real amount of data collection. We will discuss this aspect in Section~\ref{sec:pic_device_penetration}.

\begin{table*}[]
\centering
\begin{tabular}{|c|c|c|c|c|c|c|}
\hline
\textbf{} & \multicolumn{2}{c|}{\textbf{Top 100}} & \multicolumn{2}{c|}{\textbf{Top 1k}} & \multicolumn{2}{c|}{\textbf{Overall}} \\ \hline
\textbf{Rank} & Cat. & \# Domains & Cat. & \# Domains & Cat. & \# Domains \\ \hline

\textbf{1} & device info & 99 & device info & 993 & device info & 189723  \\ \hline
\textbf{2} & location info & 95 & sim card info & 891 & sim card info & 75135  \\ \hline
\textbf{3} & settings info & 92 & location info & 815 & location info & 38695  \\ \hline
\textbf{4} & email address & 87 & phone number & 646 & phone number & 17378  \\ \hline
\textbf{5} & sim card info & 85 & settings info & 593 & settings info & 15225  \\ \hline
\textbf{6} & phone number & 85 & email address & 474 & email address & 12722  \\ \hline
\textbf{7} & social network account & 68 & social network account & 280 & sms info & 3375  \\ \hline
\textbf{8} & account info & 60 & account info & 246 & social network account & 3091  \\ \hline
\textbf{9} & call log & 46 & call log & 177 & account info & 2698  \\ \hline
\textbf{10} & contact info & 40 & installed app info & 163 & contact info & 2588  \\ \hline
\textbf{11} & sms info & 38 & contact info & 138 & installed app info & 2220  \\ \hline
\textbf{12} & installed app info & 32 & sms info & 128 & call log & 1442  \\ \hline

\end{tabular}
\caption{Top 12 collected categories}
\label{tab:top_12_cats}
\end{table*}

We then investigate the diversification of private information collection by these PIC domains from a global perspective. Previous literature focused on unique hardware- and user- identifiers (UIDs) to study Advertising and Tracking Services (ATS)~\cite{razaghpanah2018apps}. It remains an open question if these PIC domains only collect UIDs given their wide presence in mobile apps. In this study, we move beyond UIDs and leverage 22 categories of private information monitored by the security company to show a holistic picture of private information collection in the mobile ecosystem. We summarize our findings in Table~\ref{tab:top_12_cats}. 
As it can be seen, the top 20 domains collect a wide spectrum of private information (\eg 14 out of 20 collect call log information, 13 out of 20 collect SMS information, etc). We can also observe that the top 10,000 PIC domains converge to collecting three types of private information - device (9,866 PIC domains), sim card (7,448 PIC domains), and location information (5,415 PIC domains) - which enable them to uniquely identify and track the end users for potential targeted advertising purpose~\cite{unni2007perceived,li2012building}. In contrast, the top 100 PIC domains focus on collecting more types of private information (\ie on average, the top 100 PIC domains collect over 8 types of private information) and build a holistic profile of users (\eg 61 out of top 100 PIC domains collect social network account information from the end users, in contrast to only 741 out of top 10,000 PIC domains collecting such information).

\noindent\textbf{Geographic differences in PIC domains.} Figure~\ref{fig:dest_dist_northamerica_by_app},~\ref{fig:dest_dist_europe_by_app} and~\ref{fig:dest_dist_asia_by_app} show the top 20 PIC domains with the largest regional app presence in North America, Europe and Asia respectively. In addition to the top global PIC domains, we uncover that certain PIC domains have a high regional app presence and were not previously reported. For example, \texttt{poseidon.mobilecore.com} (7,046 apps, 91\% of its global presence) and \texttt{seattleclouds.com} (89\% of its global presence) have high app presence in North America, Russia-based \texttt{startup.mobile.yandex.net} (1,832 apps, 72.5\% of its global presence) and \texttt{mysearch-online.com} (2,194 apps, 70\% of its global presence), respectively, have high app presence in Europe and Asia. Regarding this regional presence phenomenon, we can only speculate that it is due to the business models adopted by these companies by focusing on serving regional markets.

At the regional level, we find that the top 20 PICs contacted by apps installed on devices in different geographical regions collect different categories of private information. 
Note that we consider a PIC domain \emph{notably} collects a certain kind of private information if 20\% of apps with its presence collect such information. In North America, we observe that the top 20 PIC domains (Figure~\ref{fig:heatmap_top_20_northamerica_by_device}) mainly collect device information and sim card information, and only 3 PIC domains (\texttt{api.airpush.com}, \texttt{data.flurry.com} and \texttt{ads.mopub.com}) collect location information. In contrast, top PIC domains in Europe (Figure~\ref{fig:heatmap_top_20_europe_by_device}) and Asia (Figure~\ref{fig:heatmap_top_20_asia_by_device}) collect more diversified categories of private information. For example, 8 out of 20 top PIC domains collect location and settings information in both Europe and Asia, 4 out of top 20 PIC domains in Asia prevalently collect installed app information, with \texttt{mysearch-online.com} exclusively gathering such data.

\subsection{PIC Domains: Device Penetration Study}
\label{sec:pic_device_penetration}

In this section, we investigate the top PIC domains from the mobile device penetration rate perspective.
We show that looking at device penetration provides different results than looking at app presence only. In fact, some of the actors who manage to get their libraries installed in many apps do not manage to have a large number of users running them.

\begin{figure*}[t]
     \centering
    \begin{subfigure}[t]{0.45\textwidth}
        \includegraphics[width=\textwidth,valign=b]{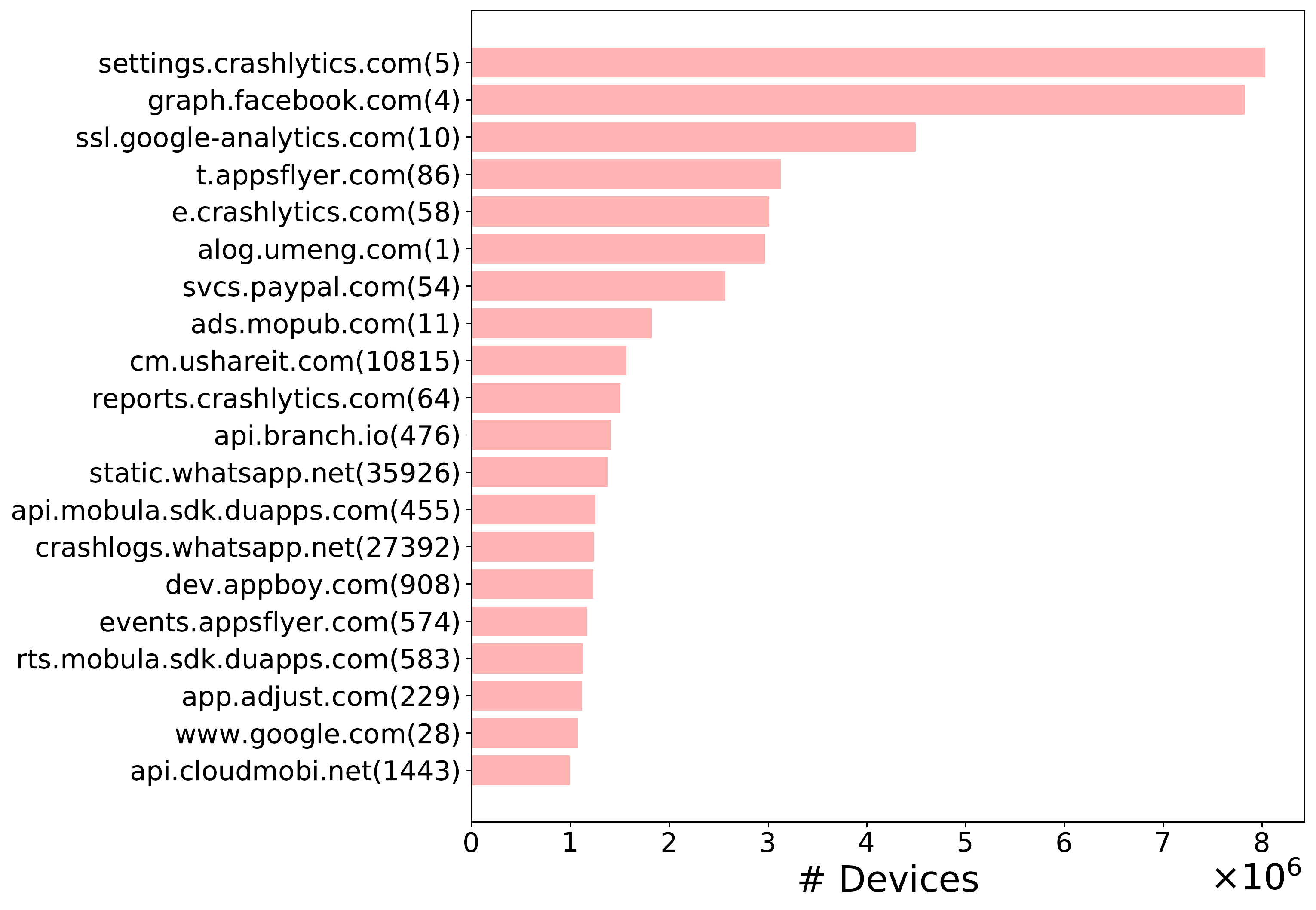}
        \caption{Global}
        \label{fig:dist_all_by_device}
    \end{subfigure}
    \hfill
    \begin{subfigure}[t]{0.45\textwidth}
        \includegraphics[width=\textwidth,valign=b]{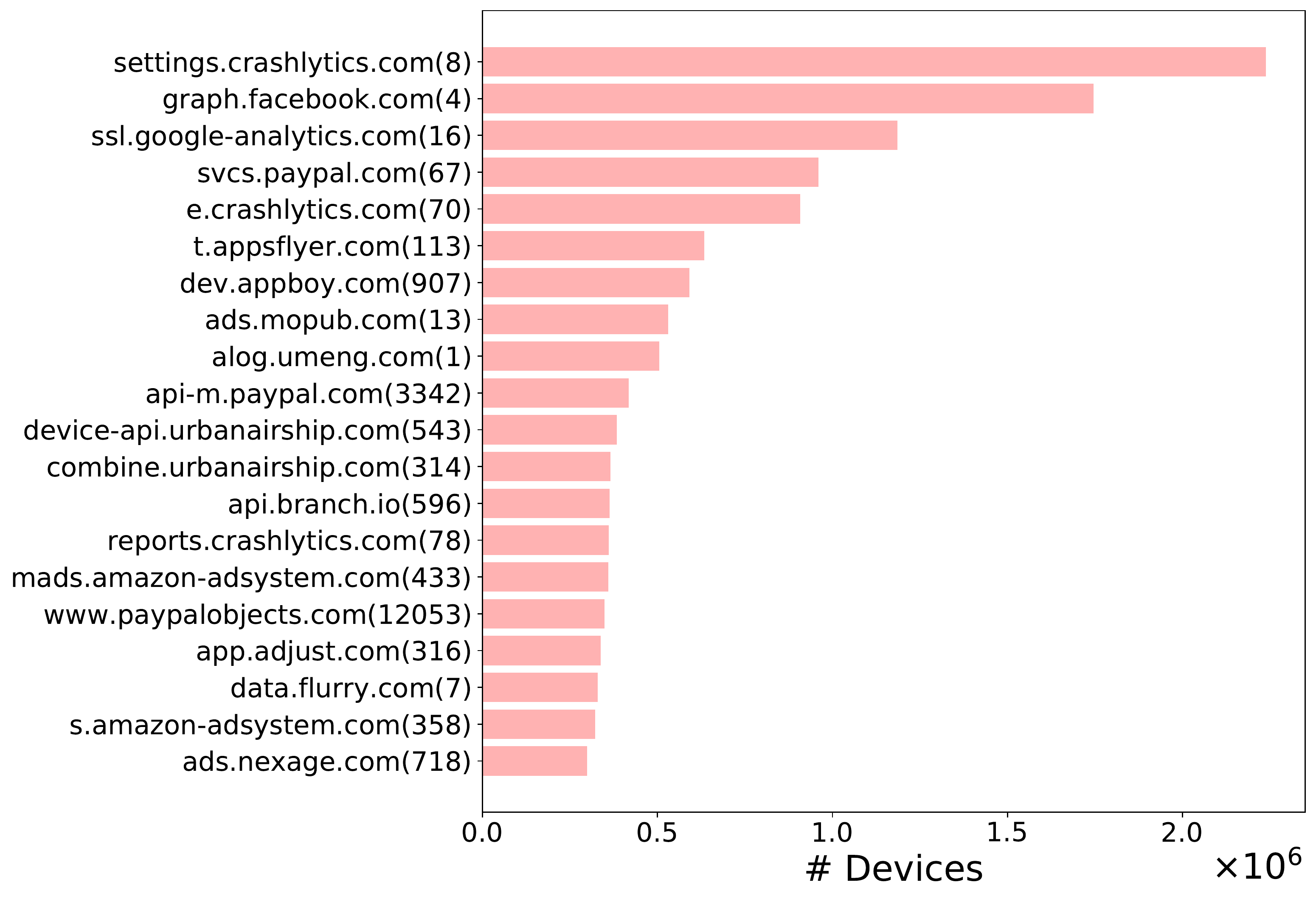}
        \caption{North America}
        \label{fig:dist_northamerica_by_device}
    \end{subfigure}
    \hfill
    \begin{subfigure}[t]{0.45\textwidth}
        \includegraphics[width=\textwidth,valign=b]{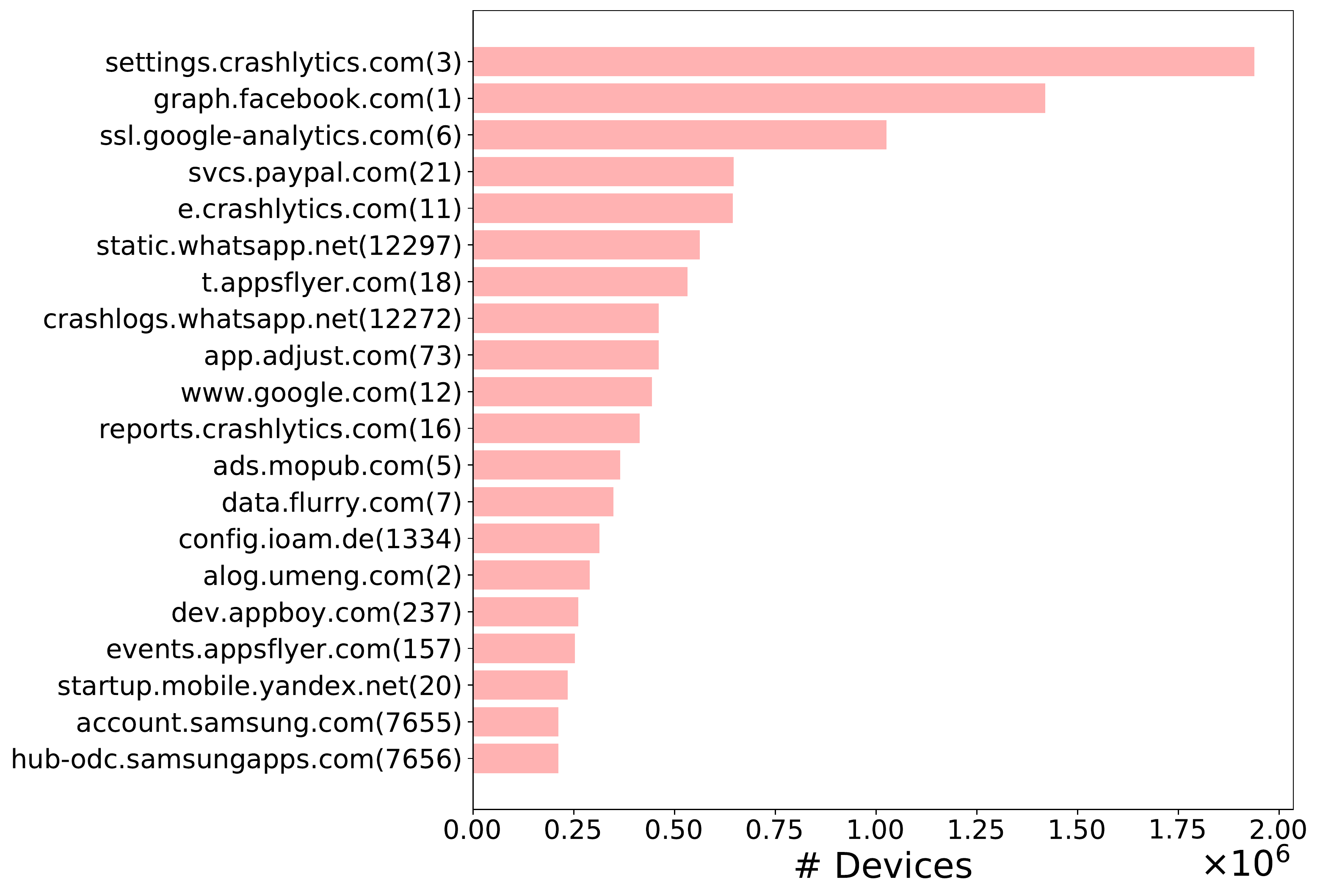}
        \caption{Europe}
        \label{fig:dist_europe_by_device}
    \end{subfigure}
    \hfill
    \begin{subfigure}[t]{0.45\textwidth}
        \includegraphics[width=\textwidth,valign=b]{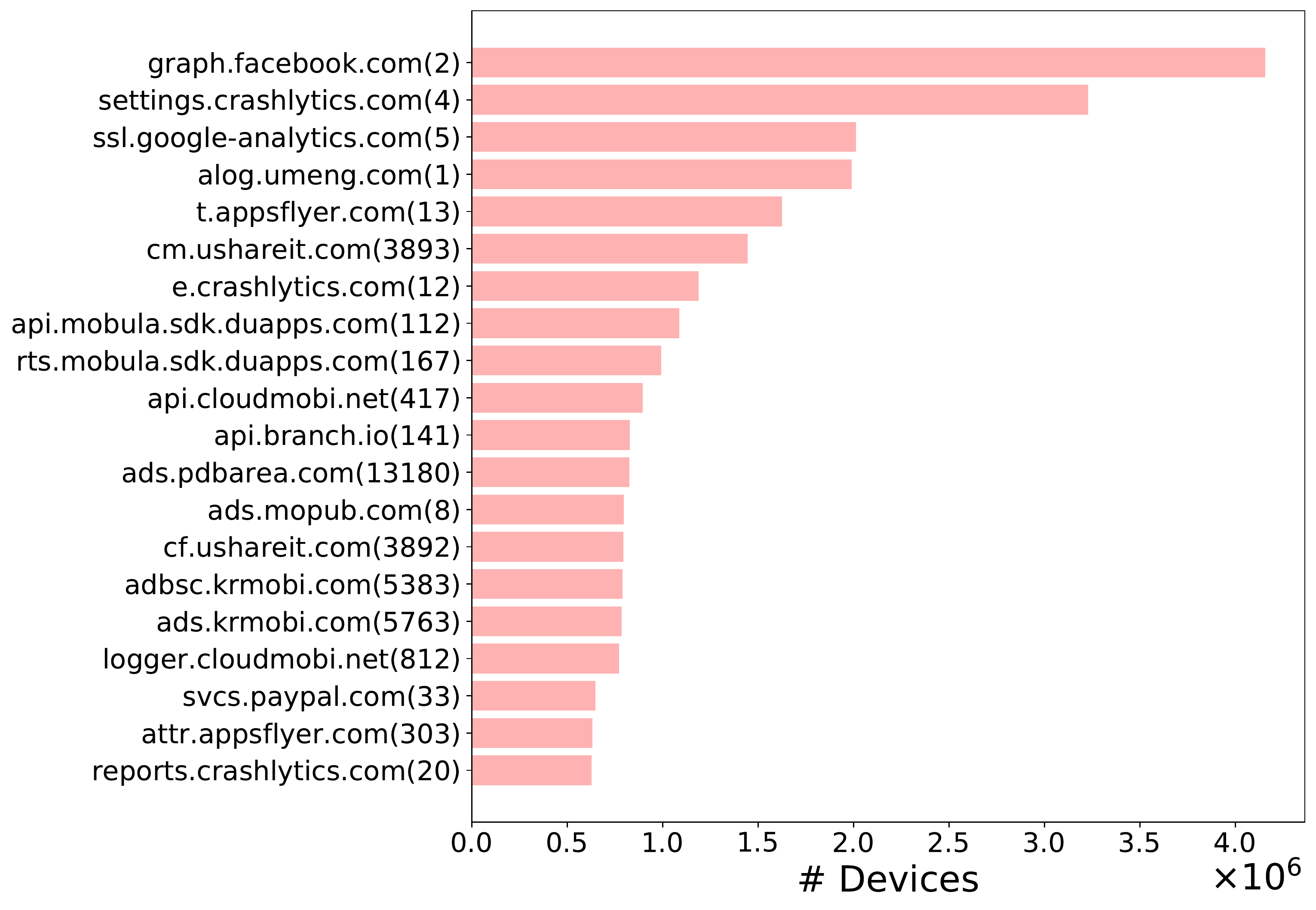}
        \caption{Asia}
        \label{fig:dist_asia_by_device}
    \end{subfigure}
    
    \caption{Top 20 PIC domains ranked by device penetration rate. The number next to a PIC domain represents its ranking by app presence.}
    \label{fig:pha_dest_dist_by_device}
\end{figure*}

\begin{figure*}[t]
     \centering
    \begin{subfigure}[t]{0.45\textwidth}
        \includegraphics[width=\textwidth,valign=b]{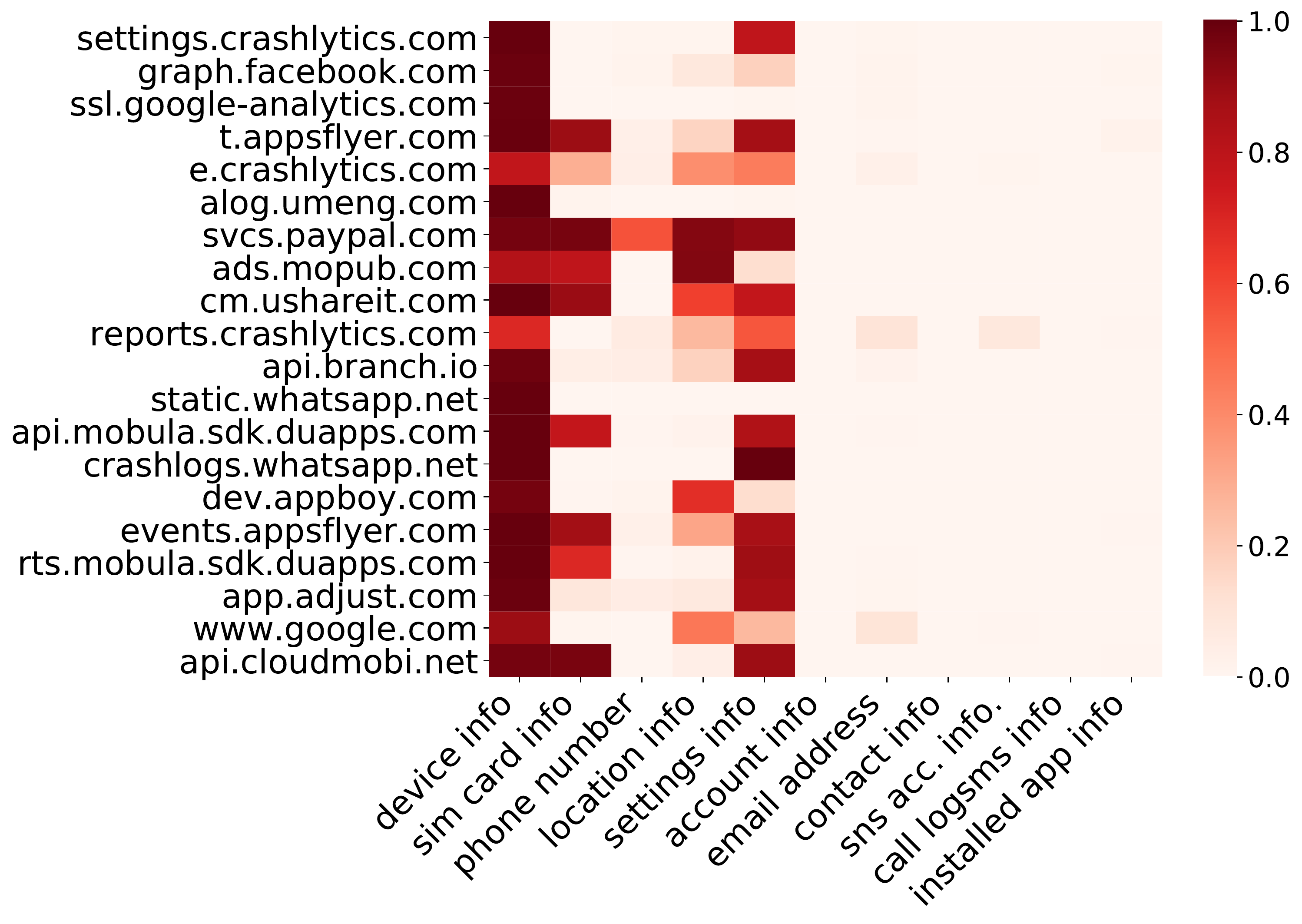}
        \caption{Global}
        \label{fig:global_dest_dist_by_device}
    \end{subfigure}
    \hfill
    \begin{subfigure}[t]{0.45\textwidth}
        \includegraphics[width=\textwidth,valign=b]{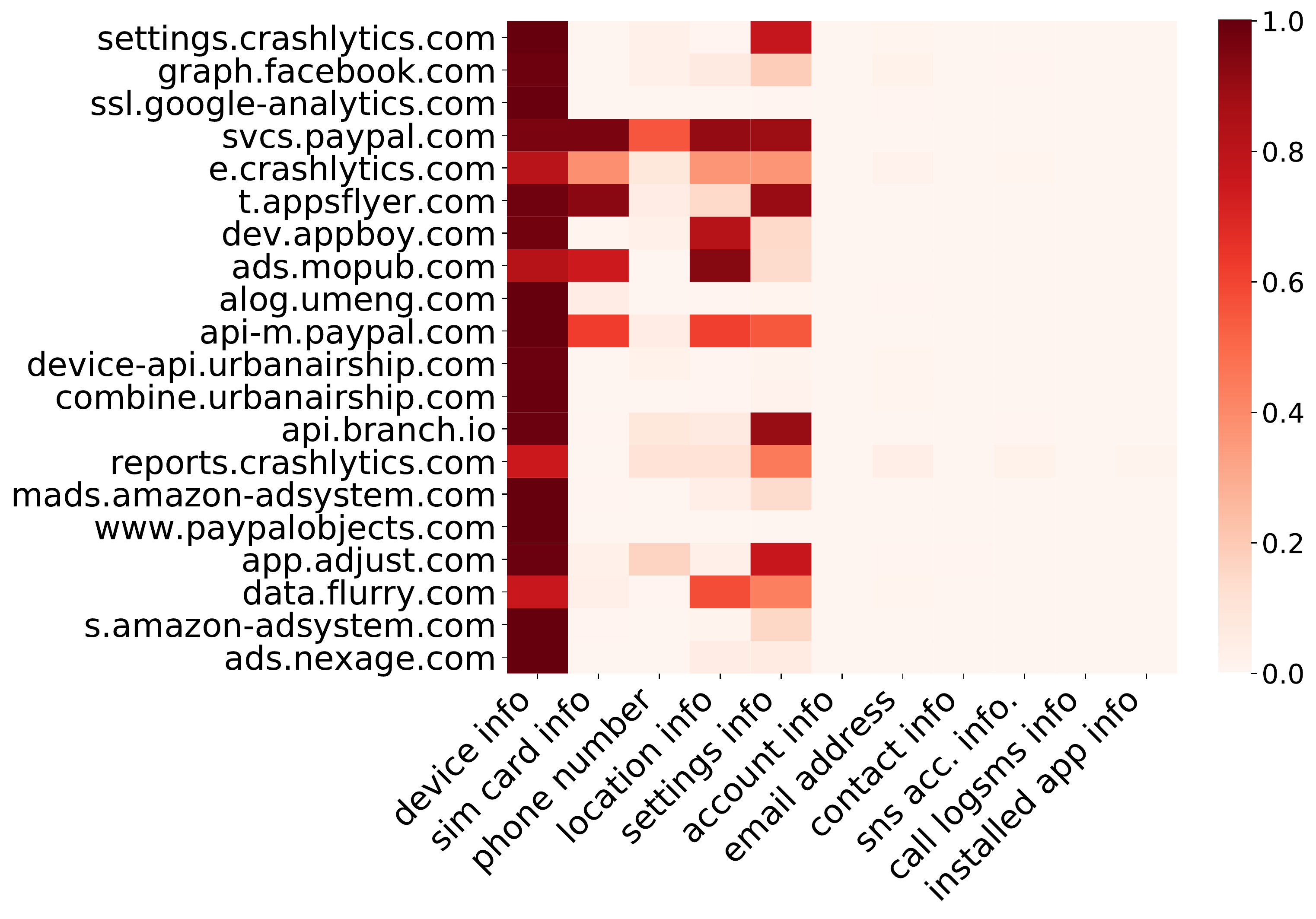}
        \caption{North America}
        \label{fig:northamerica_dest_dist_by_device}
    \end{subfigure}
    \hfill
    \begin{subfigure}[t]{0.45\textwidth}
        \includegraphics[width=\textwidth,valign=b]{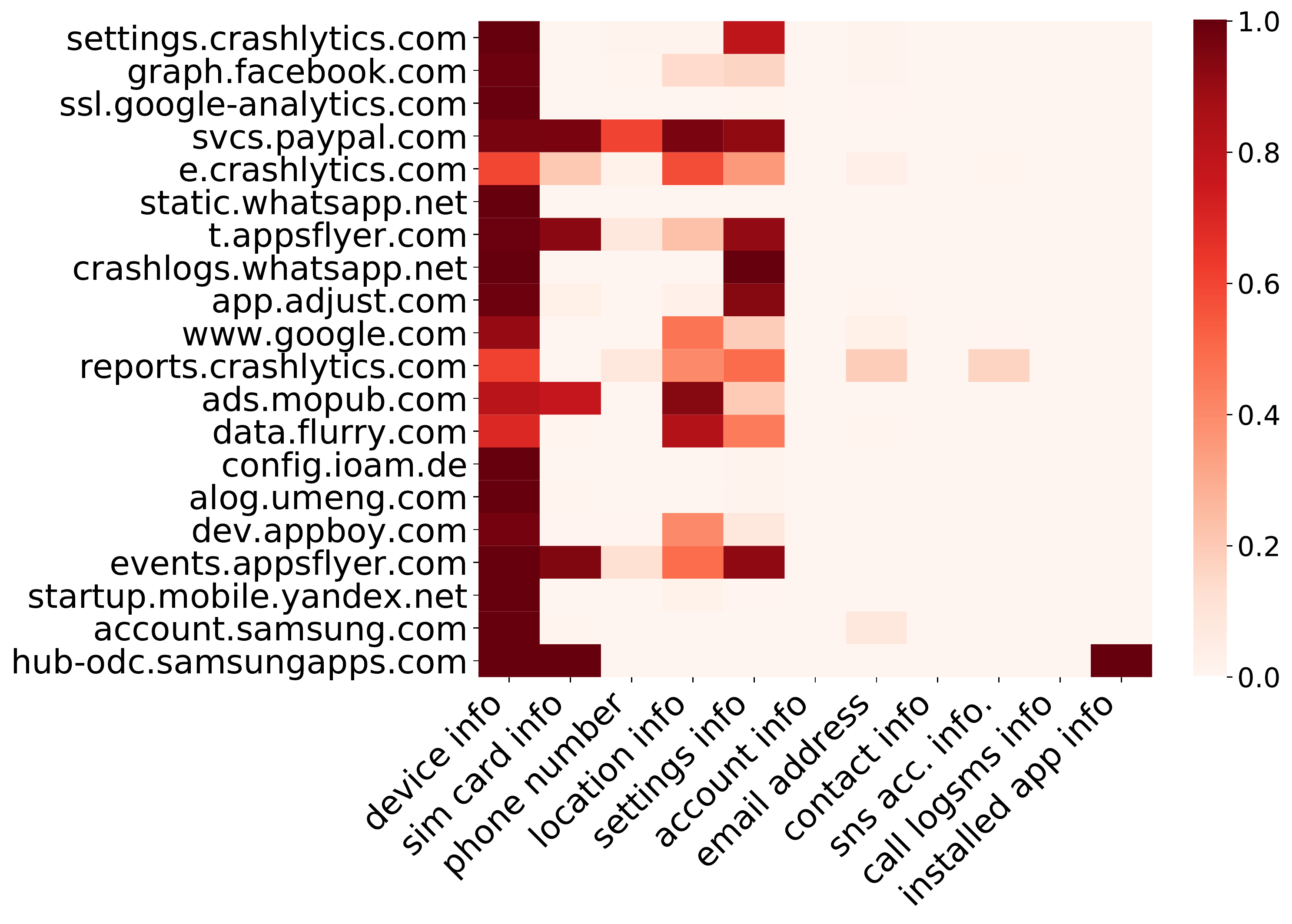}
        \caption{Europe}
        \label{fig:europe_dest_dist_by_device}
    \end{subfigure}
    \hfill
    \begin{subfigure}[t]{0.45\textwidth}
        \includegraphics[width=\textwidth,valign=b]{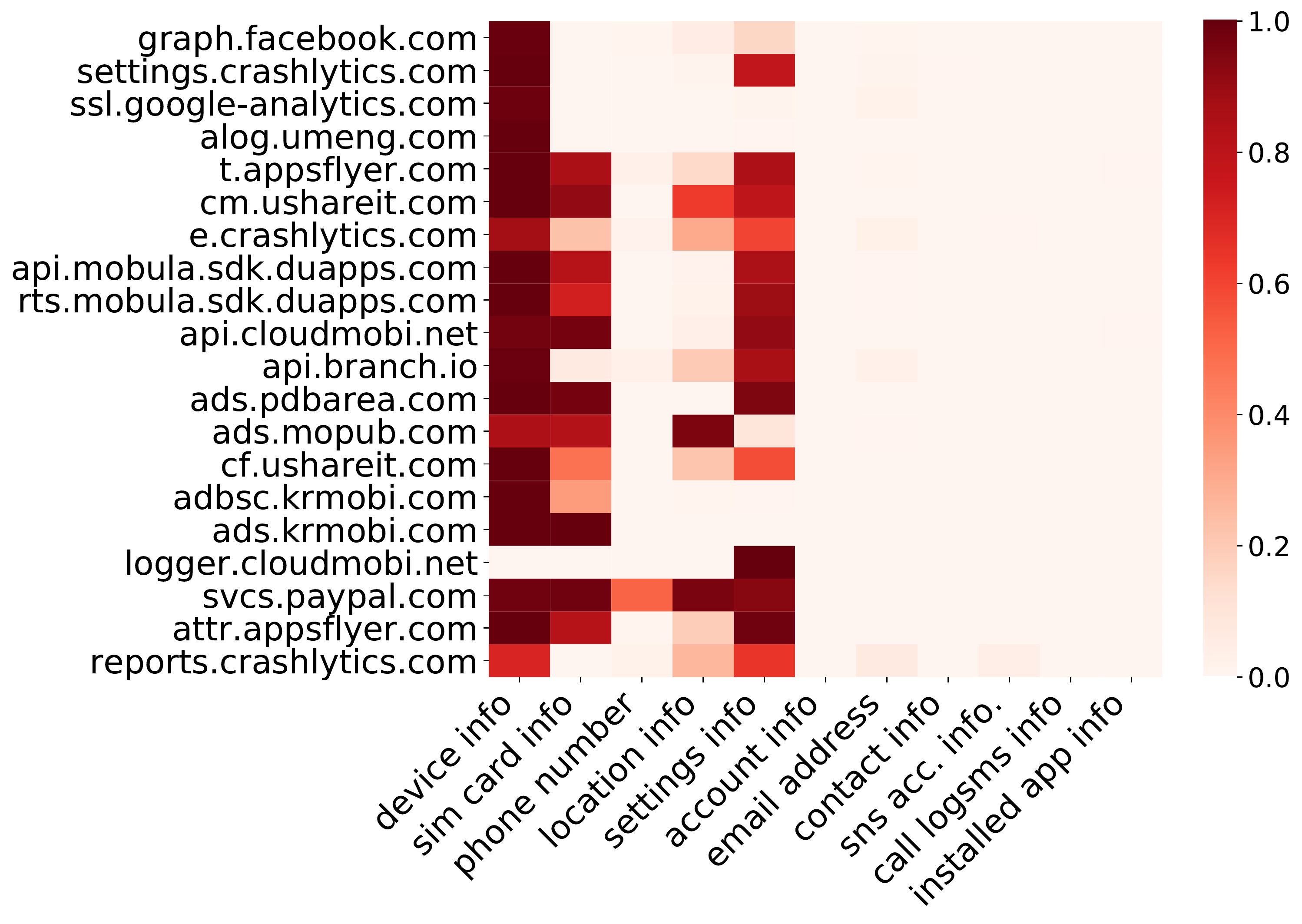}
        \caption{Asia}
        \label{fig:asia_dest_dist_by_device}
    \end{subfigure}
    \hfill
    \caption{Heatmap illustration of top 12 types of private information collected by both global and regional top 20 PIC domains. Each row is normalized to [0, 1] by a PIC domain's total device penetration rate. The darker the red implies that the more devices that a PIC domain collects information from.}
    \label{fig:dest_dist_by_device}
\end{figure*}

\noindent\textbf{Top PIC domains by device penetration rates.} In reality, a high app presence does not necessarily lead to a high device penetration rate (\ie the number of users sending information to PIC domains), whereas the latter is directly proportional to the real amount of information collection. In the rest of the section, we focus on the PIC domains that have high device penetration rates to uncover their private information collection dynamics in the real world. Figure~\ref{fig:dist_all_by_device} shows the 20 PIC domains with the largest device penetration rate globally. As we can see in Figure~\ref{fig:dist_all_by_device}, the top 3 PIC domains (\texttt{settings.crashlytics.com}, \texttt{graph.facebook.com}, and \texttt{ssl .google-analytics.com}) cover 8.03M, 7.8M, and 4.5M devices respectively, which are proportional to their app presence (see Figure~\ref{fig:dest_dist_all_by_app}). 
\texttt{alog.umeng.com}'s high app presence strategy also pays off covering roughly 3M devices.
However, quite a few PIC domains with high app presence failed to gain high device penetration rates. For example, \texttt{api.airpush.com} only covers 68K devices despite of its high app presence (21K apps, Figure~\ref{fig:dest_dist_all_by_app}). Besides, PIC domains controlled by \texttt{revmob.com}, \texttt{seattleclouds.com} and \texttt{mobilecore.com} also did not manage to have high prevalence in the devices.

\noindent\textbf{Geographic differences in PICs.} 
We also discover that different regions present different dominant PIC domains. For example, \texttt{*.urbanairship.com} and \texttt{mads.amazon-adsystem .com} have high device penetration rate in North America. \texttt{config.ioam.de} (99.4\% of its global presence) solely operates in Europe. \texttt{cm.ushareit.com} (92.4\% of its global presence), \texttt{api.mobula.sdk.duapps.com} (88\% of its global presence), \texttt{ads.pdbarea.com} (825K, 95.4\% of its global presence) and \texttt{adbsc.krmobi.com} (95.4\% of its global presence) are almost exclusively contacted by devices located in Asia. 
We further investigate the types of private information collected by PIC domains with high device penetration rates, to check if different players active in different regions are interested in different types of private information. 
Our findings are summarized in Figure~\ref{fig:dest_dist_by_device}. Each row is normalized by a PIC domain's total device penetration rate. The heatmaps illustrate the main types of information collected by the PIC domains. 
First, it is interesting to see in Figure~\ref{fig:dest_dist_by_device} that the top 20 global and regional PIC domains with high device penetration rate focus on collecting \emph{four} types of private information from the end users - \emph{device}, \emph{sim card}, \emph{location} and \emph{settings} information. For example, all of the top 20 PIC domains in Figure~\ref{fig:dest_dist_by_device} collect device information. The only exception is \texttt{logger.cloudmobi.net}, a prominent PIC active in Asia (see Figure~\ref{fig:asia_dest_dist_by_device}), which predominantly collects device setting information. Approximately 50\% of the top PIC domains collect sim card, location, and setting information at both global and regional levels. Our findings also show that certain PIC domains consistently collect multiple types of private information from devices, potentially enabling them to track the end users more systematically.  
For example, in Europe \texttt{events.appsflyer.com} (1.16M global device penetration rate)  collects device information from all devices that connected to it, and sim card information (and setting information) from 95\% of them (see Figure~\ref{fig:europe_dest_dist_by_device}).
Similarly, \texttt{ads.mopub.com} with 1.8M global device penetration rate (see Figure~\ref{fig:global_dest_dist_by_device},~\ref{fig:northamerica_dest_dist_by_device} and~\ref{fig:europe_dest_dist_by_device}) exhibits a similar behavior, \ie collects location, device, and sim card information from over 80\% of the devices that connected to it.
In Section~\ref{sec:pic_app_penetration}, we show that these two behavior patterns are different from the ones observed when looking at the top PIC domains ranked by app presence, where the intention is to collect more diversified private information (see Table~\ref{tab:top_12_cats}). Our findings can be treated as the profiles of PIC domains, and help the community understand their behavior in fine granularity (\eg understanding the correlation between domain naming conventions and the types of private information collected).

\noindent\textbf{Summary of findings.} We found that looking at app presence can provide misleading results. In fact, some of the actors who managed to get their libraries installed in many apps failed to have many users running them. Further information can be found in Section~\ref{sec:pic_app_penetration}. We also found that certain PIC domains consistently collect multiple types of private information from the devices and are capable of tracking the end users more systematically.   
We observed different regional players targeting users in different continents, and collecting different types of private information.
Following these observations, we will further discuss the data controllers behind these PIC domains and the implications of data protection in Section~\ref{sec:data_controller}.

\section{Private Information Destinations}
\label{sec:data_flow}

In the previous section, we focused on end user devices, looking at the top PIC domains that collected private information from them. 
In this section, we focus on the destination of those private information flows, aiming to understand the countries where these flows terminate. 

\begin{figure}
\centering
\includegraphics[width=0.9\linewidth]{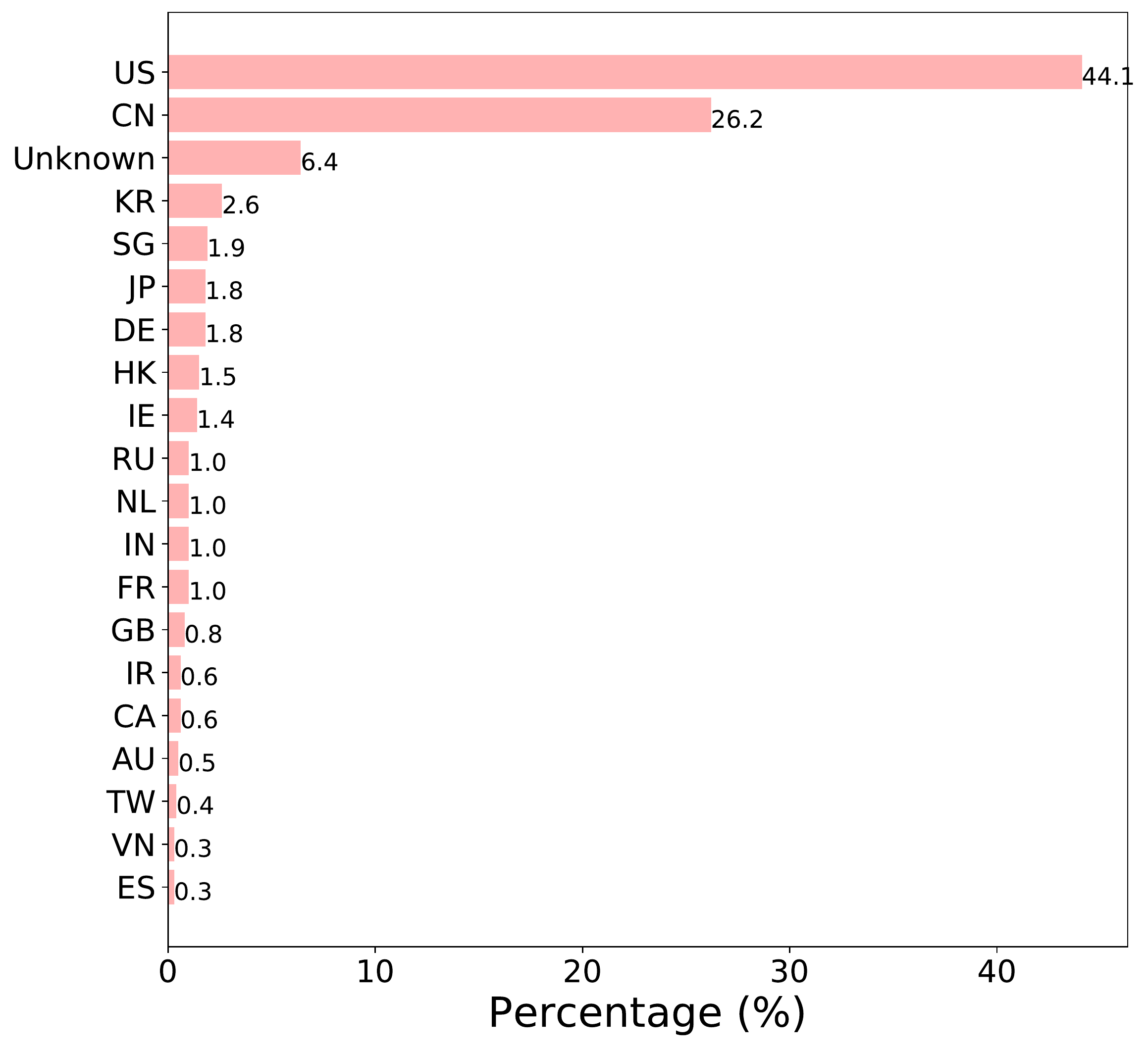}
\caption{Global top 20 countries ranked by the number of PIC domains hosted. }
\label{fig:top_20_ats_all_by_device}
\end{figure}

\noindent \textbf{Geolocation of PIC Domains.} We leverage the technique detailed in Section~\ref{sec:datasets} to uncover the geolocation of the PIC domains and summarize our findings in Figure~\ref{fig:top_20_ats_all_by_device}.
Our analysis reveals that United State and China are the largest two countries hosting the PIC domains. The United States hosts 44\% of PIC domains, which is in line with the previous literature~\cite{razaghpanah2018apps} and China
hosts 26.1\% of PIC domains. 
This figure is three times higher than previously reported~\cite{razaghpanah2018apps}. 
PIC domains hosted in the US and China collect private information from 14M devices (80.9\% of global devices) and 4.6M devices (26.5\% of global devices) respectively.  
Other countries host significantly fewer PIC domains compared to the United States and China (\eg South Korea, ranked 3rd in the list, hosts merely 2.6\% of the PIC domains). Note that the geolocation of 6.4\% of PIC domains could not be identified because our approach cannot trace their historical domain records. 

\begin{figure*}
     \centering
    \begin{subfigure}[t]{0.32\textwidth}
        \includegraphics[width=\textwidth,valign=b]{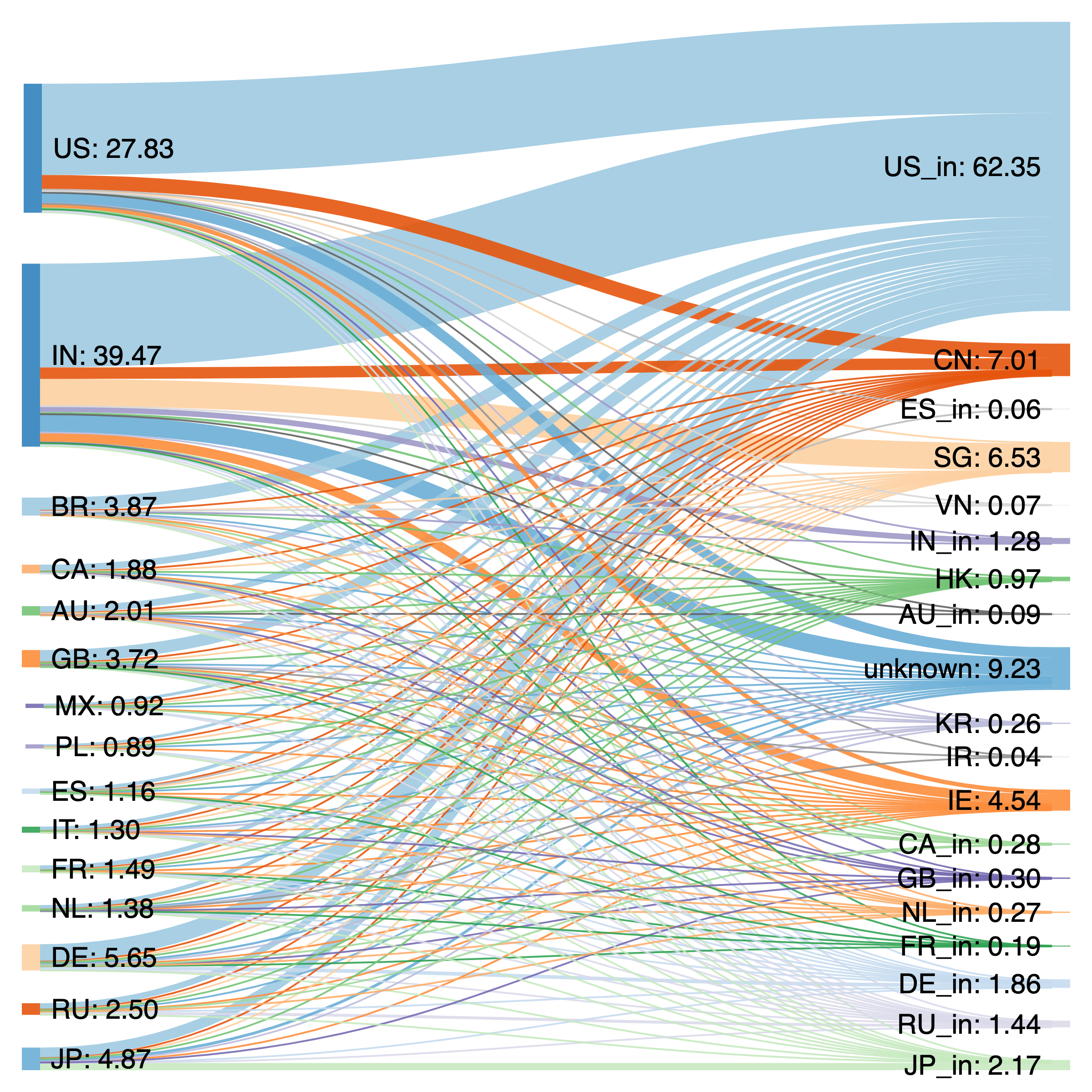}
        \caption{}
        \label{fig:global_data_flow}
    \end{subfigure}
    \hfill
    \begin{subfigure}[t]{0.32\textwidth}
        \includegraphics[width=\textwidth,valign=b]{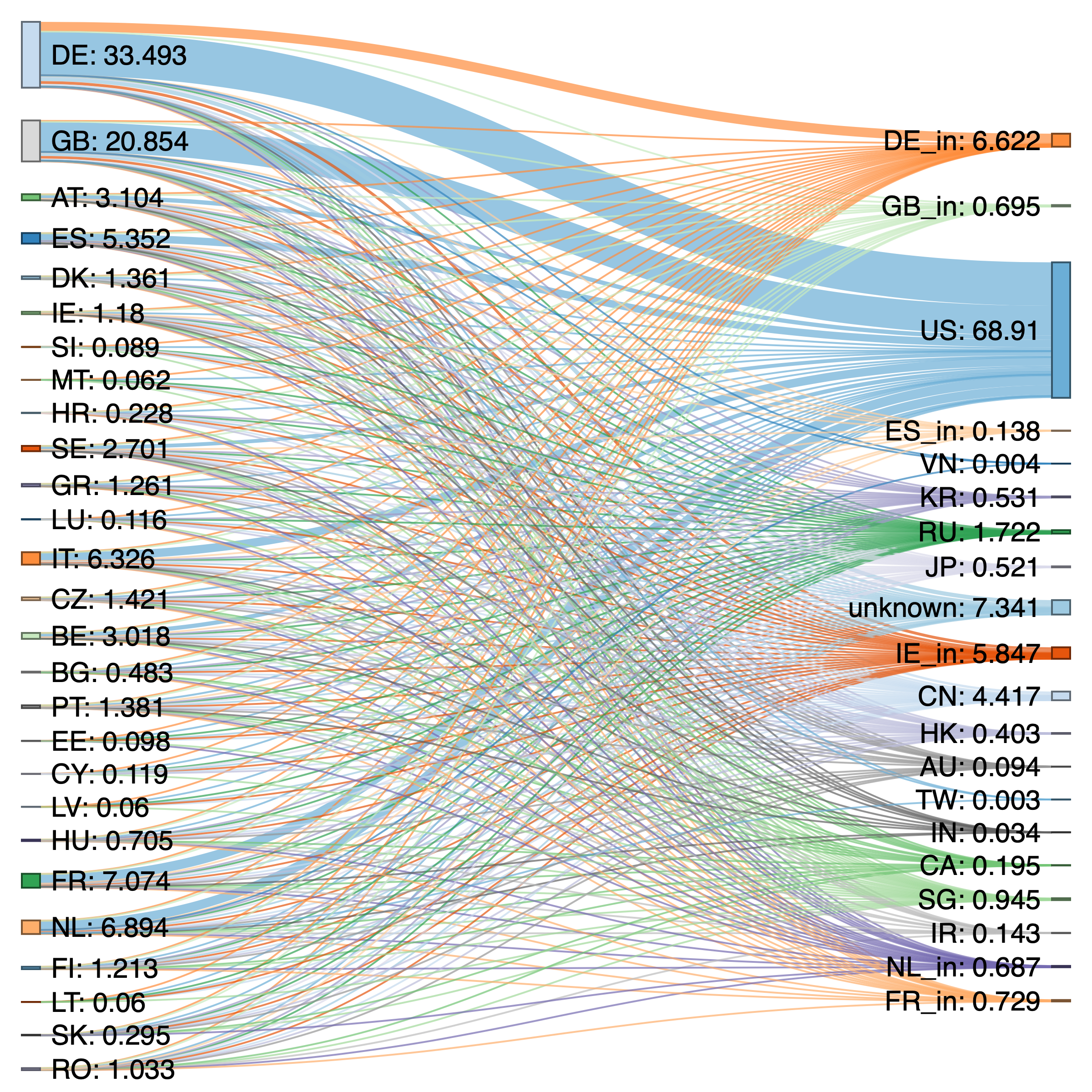}
        \caption{}
        \label{fig:before_gdpr_data_flow}
    \end{subfigure}
    \hfill
    \begin{subfigure}[t]{0.32\textwidth}
        \includegraphics[width=\textwidth,valign=b]{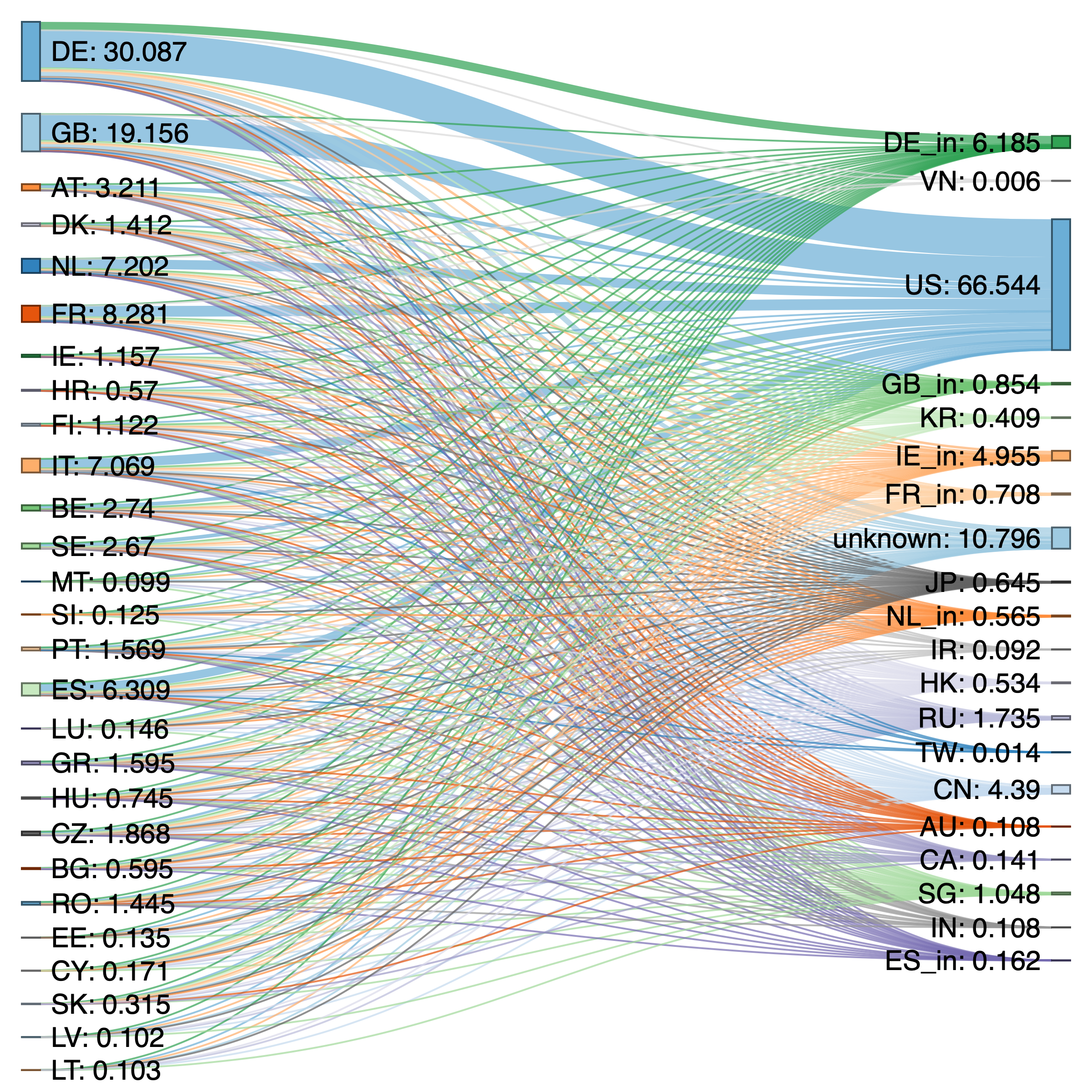}
        \caption{}
        \label{fig:after_gdpr_data_flow}
    \end{subfigure}
    \hfill
    \caption{Sankey diagrams illustrating 1) Global private information flows between the top 15 countries (ranked by the number of devices) and top 20 PIC domain locations (a) and 2) Private information flows between EU28 and top 20 PIC domain locations before (b) and after (c) GDPR. Note that the left side of the diagrams represents the origin of information flows and the right side represents where the information flows terminate. We add a postfix `\_in' to the country code at the right hand side in case of private information flows originating and terminating at the same country.}
    \label{fig:data_flow}

\end{figure*}

\noindent \textbf{Global private information flow}. As we saw in Section~\ref{sec:pic_landscape}, a PIC domain can collect multiple types of private information from the end user. 
We further investigate the global private information flow from the mobile devices to the PIC domains. The result is shown in Figure~\ref{fig:global_data_flow}. PIC domains hosted in the United States collect 62\% (of which 42.3\% coming from out of the country) of global private information flows. PIC domains hosted in China
collect 7\% of private information flows from 4.59M devices globally. This figure is almost four times more than previously reported~\cite{razaghpanah2018apps}. At the same time, PIC domains hosted in Singapore collect 6.53\% of global private information flows (mainly from India). The rest of the countries shown in Figure~\ref{fig:global_data_flow} notably collect much less private information comparing to these three countries.

\noindent \textbf{European private information flow and the effect of GDPR~\cite{voigt2017eu}.} The European Union's (EU) General Data Protection Regulation (GDPR) entered into effect on May 25th, 2018. It imposes obligations onto organizations in any country so long as they target or collect data related to people in EU countries (EU28).
If data is being transferred to a third-party and/or outside the EU28, GDPR requires that data subjects must be clearly informed about the extent of data collection, the legal basis for the processing of personal data, how long data is retained. In light of this legislation, we measure the private information flows originated from EU countries before (January 5th, 2018 - May 24th, 2018) and after (May 26th, 2018 - September 30th, 2019) the GDPR effective date, and check if GDPR has a real-world impact to private information collection. Our findings are shown in Figure~\ref{fig:before_gdpr_data_flow} and~\ref{fig:after_gdpr_data_flow}. As we can see, private information confinement within the EU is low. PIC domains hosted in the United States dominate the private information collection in the EU, collecting 68\% and 66\% of European private information flows respectively before and after the GDPR. This figure is 30\% lower than previously reported 89.2\%~\cite{razaghpanah2018apps}. 
At the same time, Germany and Ireland are the only two European countries that host a reasonable portion of PIC domains and good control of private information can be applied, while the other European countries hosting a very small fraction of PIC domains and US remains the largest hosting country. Notably, we uncover that approximately 4.4\% and 1.7\% of private information flows are collected by PIC domains hosted in Russia and China respectively~\cite{zhao2019data,zhao2016protecting}.

It is also interesting to see that private information collection in Europe is not affected by GDPR in general. As we can see in Figure~\ref{fig:before_gdpr_data_flow} and Figure~\ref{fig:after_gdpr_data_flow}, the fractions of private information collected by these PIC domains (and consequently the countries hosting them) remains stable regardless of the implementation of GDPR. Our results show that GDPR has not stopped companies from collecting private information from end users as long as their services are GDPR-compliant, partially because that the GDPR treats first-party data uses more leniently~\cite{hoofnagle2019european}. However, it remains an unanswered question, especially to the consumers, how to trace their private information after sharing with the GDPR-compliant companies, and how accountability can be truly guaranteed~\cite{voss2019personal,minssen2020eu,mulder2019privacy}. For instance, which company should be held accountable if a device identifier was abused (\eg targeted advertising) while the majority of apps in mobile devices collect device identification information as shown in Section~\ref{sec:pic_device_penetration}? We aim at studying this question in Section~\ref{sec:data_controller}.

\begin{figure}[t]
\centering
\includegraphics[width=0.9\linewidth]{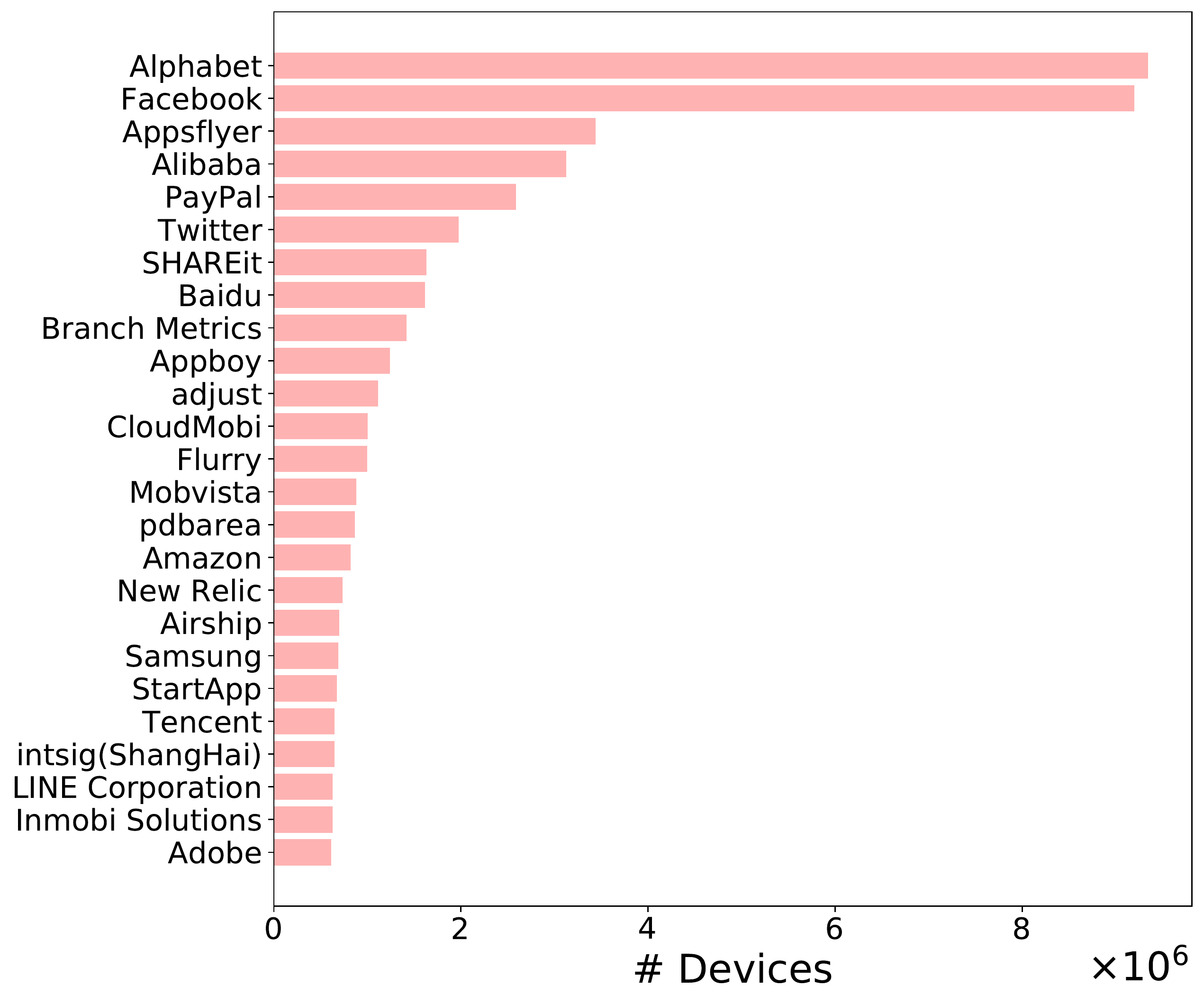}
\caption{Global top 25 data controllers ranked by the fraction of devices they collect private information from. These 25 data controllers collect private information from a total of 13.9M devices covering 80.2\% of all devices used in this study.}
\label{fig:top_25_companies}
\end{figure}

\section{Data Processors and Controllers}
\label{sec:data_controller}

In the previous sections, we provided an overview of the landscape of private data collection from mobile devices (Section~\ref{sec:pic_landscape}) and of the countries where private information is sent to (Section~\ref{sec:data_flow}). In this section, we aim at understanding the characteristics of the data processors and controllers who ultimately obtain and process the private information and the implications of their privacy policies to the end users.

\noindent \textbf{Overview of top data processors and controllers.} We select the top 10k PIC domains covering all the devices in this study, and use the technique detailed in Section~\ref{sec:datasets} to uncover the ownership of the PIC domains. The top 25 data processors and controllers (ranked by the fraction of devices they collect private information from) are shown in Figure~\ref{fig:top_25_companies}. In total, these 25 data processors and controllers collect private information from 13.9M devices  (80.2\% of all devices used in this study). Facebook and Alphabet are the two dominant data controllers, collecting private information from 9.3M and 9.1M devices respectively. AppsFlyer is the third largest data processor/controller collecting information from 3.4M devices. It is worth noting that there are six Chinese companies among the global top 25 data processors and controllers: Alibaba (3.1M), Baidu (1.6M), CloudMobi (1.0M), MobVista (880K), Tencent (650K), and Intsig(Shanghai) (646K). In total, these six companies are collecting private information from 4.55M devices (\ie 26\% of total devices). 

\begin{figure}
\centering
\includegraphics[width=0.9\linewidth]{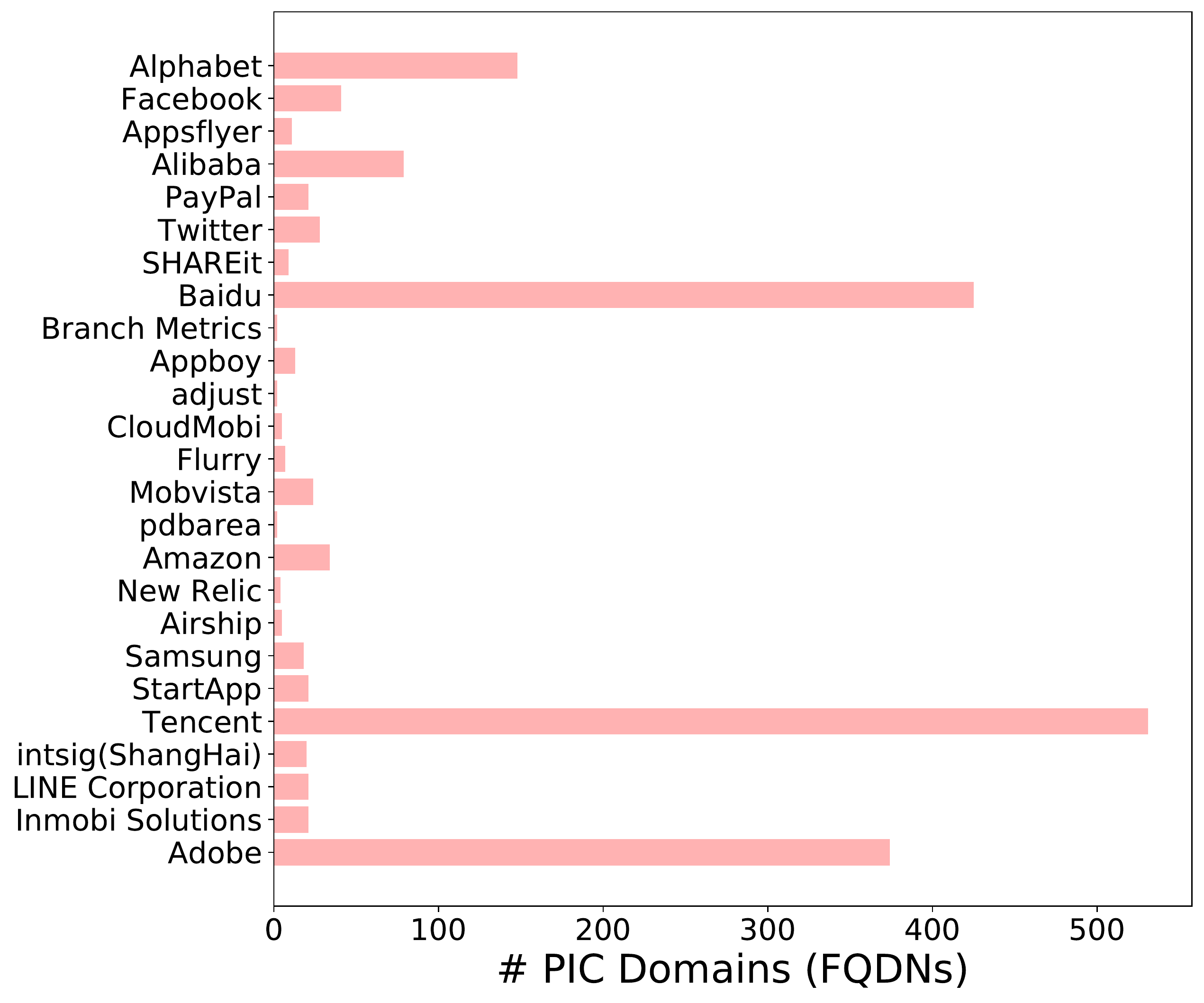}
\caption{Domain distribution: top 25 data controllers.}
\label{fig:domain_distribution_top_25_companies}
\end{figure}

\noindent \textbf{Operation of top data processors and controllers.} 
We analyze the domain distribution of these top 25 data processors and controllers to understand more details on their infrastructure and on their operational strategies. 
Our findings are summarized in Figure~\ref{fig:domain_distribution_top_25_companies}.
In total, 16 out of the top 25 data processors and controllers have no more than 21 PIC domains. For example, AppsFlyer, the third largest data processor/controller, has only 11 PIC domains in our dataset. It is evident that the majority of the data controllers prefer to control data flow via several API gateways. At the same time, Baidu (425 PIC domains), Tencent (531 PIC domains), and Adobe (374) prefer to use many loosely coupled services to collect data since their operational strategies rely on the Cloud infrastructure. For example, DU Ad Platform (\texttt{*.duapp.com}, part of Baidu) almost exclusively runs in the AWS infrastructure, QQ platform (\texttt{*.qq.com}, part of Tencent) operates in Tencent owned Cloud infrastructure, and 2o7 (\texttt{*.2o7.net}) is part of Adobe Marketing Cloud. Note that previous literature~\cite{razaghpanah2018apps} found that ``\emph{292 parent organizations that own nearly 2,000 ATS and ATS-C domains.}'' Our findings, however, indicate that these data controllers may own more PIC domains that previously thought.

\begin{figure}
\centering
\includegraphics[width=0.9\linewidth]{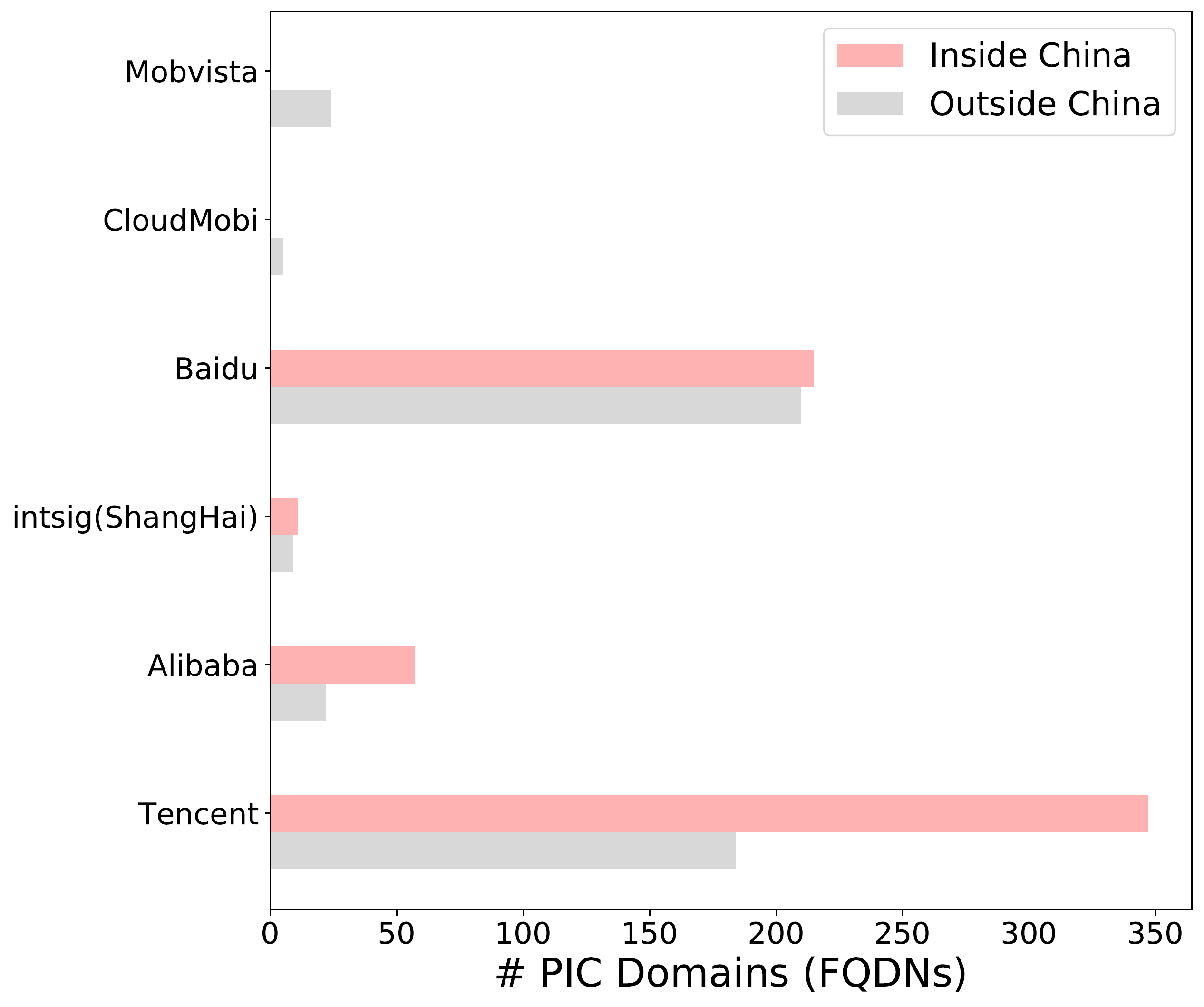}
\caption{Domain distribution: top 6 Chinese data controllers.}
\label{fig:domain_by_chinese_controllers}
\end{figure}

\noindent \textbf{Cross-border transfer, non-EU data processor and controllers, and implications to data protection.} Based on our factual findings, we use Chinese companies as a case study to quantitatively and objectively understand the implications of users' private information collection when involving cross-border data transfer~\cite{voss2019personal,minssen2020eu,mulder2019privacy,ITA2020} and how it becomes more difficult to trace how this data flows. As mentioned before, the top six Chinese companies are collecting private information from 4.55M devices. Superficially, such coverage seems in line with our findings in Section~\ref{sec:data_flow} where we found that 7\% of private information flows from 4.59M devices globally flow to China. We further investigate the geolocation of the PIC domains controlled by these companies and see if these domains are hosted in China using the technique detailed in Section~\ref{sec:datasets}. 

For example, Baidu has 210 PIC domains hosted outside China, mainly because the \texttt{*.duapp.com} PIC domains (owned by its subsidized DU Ad Platform) are hosted in AWS (USA). Besides, \texttt{*.mobvista.com} is hosted in Amazon Web Services (AWS) and \texttt{*.cloudmobi.net}, has a mixture of hosting environments in the US and Singapore. 
We report more details about the country distribution of Chinese data controllers in Figure~\ref{fig:domain_by_chinese_controllers}. The figure confirms that many PIC domains owned by these companies are not hosted in China.
Such operational strategy employed by these data controllers, however, leads to undesirable implications for data protection. For example, the DU Ad Platform (partnering with Facebook, Alphabet, appnext, etc.) states in its privacy policy\footnote{http://ad.duapps.com/gdpr/index.html\#title-2} that personal information could be ``\emph{shared with any organization part of Baidu Group}'' and ``\emph{may be transferred to countries which provide an adequate level of protection}.'' In this case, even though private information flows terminate in the AWS Cloud, such data could still be transferred to third countries. Moreover, Mobvista (partnering with Baidu, TikTok, etc) explicitly claims in its privacy policy\footnote{https://www.mobvista.com/en/privacy/} that private information would be ``\emph{transferred to recipients in countries located outside the EEA (including in Singapore where the Site is hosted) which do not provide a similar or adequate level of protection to that provided by countries in the EEA}.''We acknowledge that without knowing more about the actual underlying contractual relationships it is difficult to draw conclusions on how data is further processed by those entities. Nevertheless, it remains an open yet important question on how to protect and audit the usage of such data flows terminated at the PIC domains owned by these companies with data transfers to third countries explicitly stated in the privacy policy. We hope that our findings will motivate lawmakers to consider how to address such issues in the future legislations, and more importantly, encourage the commercial partners of these companies to design rigorous policies to protect user private information when sharing data cross-border.

\noindent\textbf{Summary of findings.} We found that the top 25 data processors and controllers  can collect private information from an overwhelming 80.2\% of all devices. 6 top Chinese data controllers provided privacy policies and hosted part of their infrastructure in countries with rigorous data protection laws. However, they also allow data transfer to third countries and may incur technical and legal complications on how to further protect private information~\cite{zhao2019data,zhao2016protecting,voss2019personal,minssen2020eu,mulder2019privacy}.

\section{Characterization of PHA Private Information Collection}
\label{sec:pha_data_flow}

Potentially harmful applications (PHAs)\cite{googlereport} are apps that could put users, user data, or devices at risk (\eg trojan, spyware, etc.). Some of them aren’t strictly malware but are harmful to the software ecosystem (\eg impersonating other apps). These PHAs have been substantially discussed and studied in the previous literature~\cite{lindorfer2014andradar,zhou2012dissecting,felt2011survey,faruki2014android,chatterjee2018spyware}. 
In this section, we focus on understanding what private information is collected by PHAs.
In particular, we aim to understand whether the type of information collected by PHAs is different from the one collected by regular apps.

\begin{table}
    \centering
    \resizebox{0.7\linewidth}{!} {
    \begin{tabular}{llll}
    \toprule
    Rank & Cat. & \# PHAs  & \# Dev.\\
    \midrule
    1 & Device Info & 295K & 1.45M \\
    2 & Sim Card Info & 167K & 993K \\
    3 & Location Info & 127K & 670K \\
    4 & \textbf{Operator Info} & \textbf{116K} & \textbf{393K} \\
    5 & Installed App Info & 91K & 486K \\
    6 & Phone Number & 75K & 364K \\
    7 & \textbf{Running App Info} & \textbf{63K} & \textbf{280K} \\
    8 & Account Info & 17K & 73K \\
    9 & Settings Info & 10K & 376K \\
    10 & Email Address & 4K & 107K \\
    \bottomrule
    \end{tabular}
    }
    \caption{Top 10 private information collected by PHAs on a global scale.}
    \label{tab:top_10_PI_by_PHAs}
\end{table}

\begin{figure}
\centering
\includegraphics[width=\linewidth]{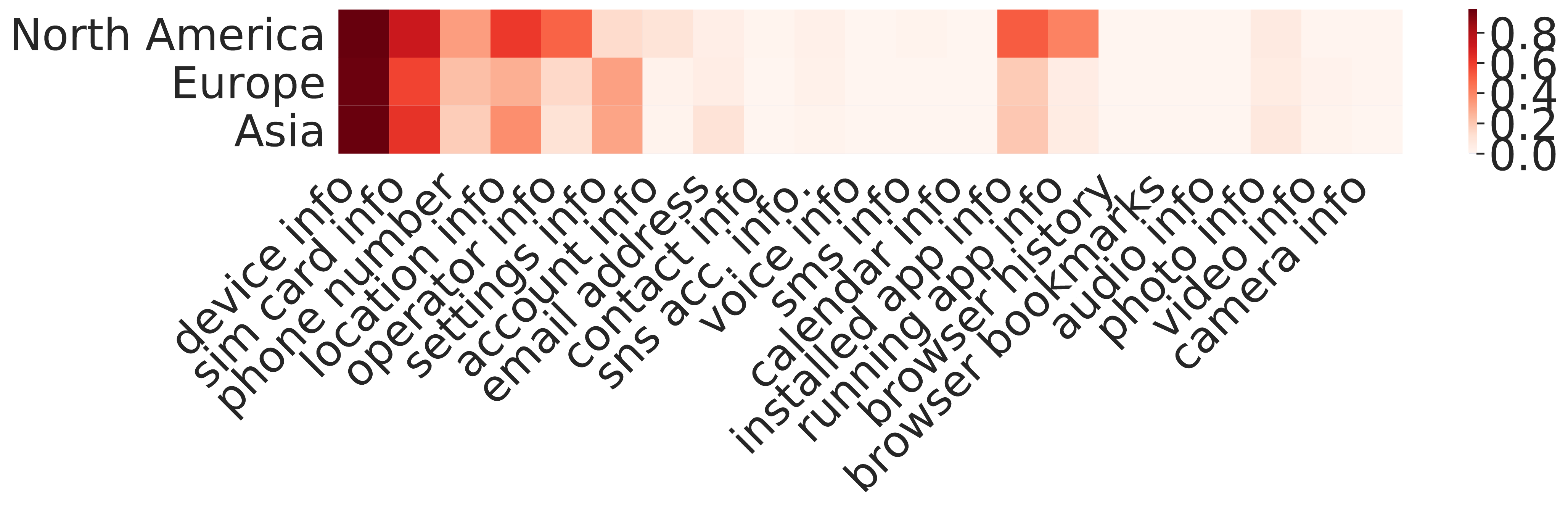}
\caption{Heatmap illustration of regional private information collection by PHAs.}
\label{fig:pic_pha}
\end{figure}

\noindent \textbf{Private information collection by PHAs.} We consider a SHA2 as potentially harmful if it is flagged by at least 6 AV companies in VirusTotal (see Section~\ref{sec:datasets}). Together with mobile app reputation data, we identify 3.5M SHA2s associated with 1.2M unique PHA app names that were installed on 3.8M devices. Following the analytical process used in Section~\ref{sec:pic_landscape}, we uncover the top 10 types of private information collected by PHAs and summarize our findings in Table~\ref{tab:top_10_PI_by_PHAs}. We can see that PHAs mainly collect tracking information, \eg device info, sim card, location, etc. Besides, 116K PHAs (covering 393K devices) collect operator information and 63K PHAs (covering 280K devices) also collect running app information on a global scale. This is more aggressive comparing to the private information collection behavior comparing to 43K/42K benign apps respectively collecting such information. As we can see in Figure~\ref{fig:pic_pha}, the majority of these aggressive PHAs are installed on devices in North America. Note that such aggressive private information collection behavior enables adversaries to better profile end users and may lead to some intrusive monetization actions. For example, we uncover that 590K devices with PHAs presence are affected by notification bar ads (\ie ads are displayed as app notifications) and 317k devices suffer from short-cut ads (\ie targeted ads are placed on the home screen). Yet, only 230K devices with PHA installations exhibit in-context ads behavior (\ie normal behavior as ads are displayed inside an app). However, due to the limitation of our system, we are not able to measure the content correlation between private information collected by PHAs and the subject of advertisement displayed as shortcuts on the devices. We also identify 1,549 PHAs (4,930 SHA2s) that read/sent SMS from 4,461 devices. Even though such SMS leakage is minor in terms of device prevalence ratio, in light of the recent discussion of limitation of SMS-based 2FA authentication\footnote{https://krebsonsecurity.com/2018/08/reddit-breach-highlights-limits-of-sms-based-authentication/}, our findings show that the possibility of such breaches still exists in the wild.

\begin{figure}
\centering
\includegraphics[width=0.9\linewidth]{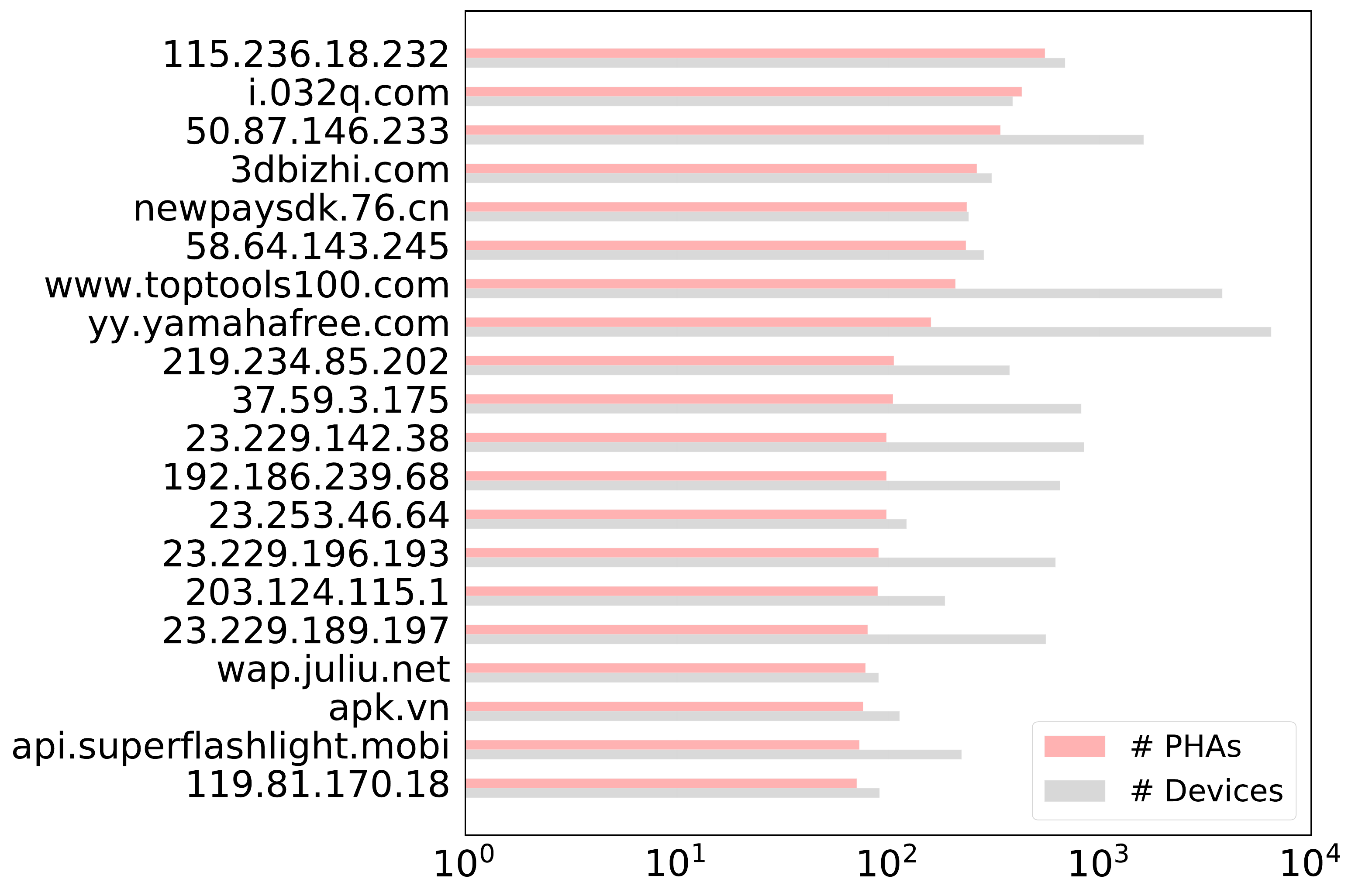}
\caption{\textbf{(log-scale)} Top 20 malicious IPs/Domains ranked by the number of PHAs.}
\label{fig:top_20_malicious_domains}
\end{figure}

\noindent \textbf{Communications with malicious domains.} We compile a blacklist of IPs/domains that have been involved with malicious activities from various sources (see Section~\ref{sec:datasets}), and aim at understanding if PHAs send private information collected from these devices to malicious domains. Figure~\ref{fig:top_20_malicious_domains} shows the 20 malicious domains with the largest app presence and the fraction of devices connecting to them. As we can see in Figure~\ref{fig:top_20_malicious_domains}, \texttt{115.236.18.232} has the largest app presence and was contacted by 550 PHAs collecting data from 686 devices. \texttt{www.toptools100.com} and \texttt{yy.yamahafree.com} have  higher device penetration rates, respectively showing communications with 3,789 and 6,455 devices respectively. In general, we find that only a small portion of PHAs communicate with known malicious hosts and domains, and such domains have limited device coverage. This is different from PC malware while a considerable fraction of malware connect with malicious domains and are part of botnets~\cite{antonakakis2011detecting,plohmann2016comprehensive,ife2019waves}.

\noindent \textbf{Summary of findings.} We found that PHAs are more aggressive comparing to generic private information collection behavior, leading to intrusive monetization actions. However, communications with malicious domains are less pervasive comparing to desktop applications.

\section{Discussion and Limitations}
\label{sec:discussion}

\noindent \textbf{Implications for the research community.} 
Our study shows that looking at device penetration provides different results than looking at apps only. 
Designing measurement studies focused on executing apps could lead to conclusions that are biased and do not reflect real malicious activity in the wild.
In fact, some of the actors who manage to get their libraries installed in many apps do not manage to have many users running them. 
In light of this, we hope that our study can inspire security researchers to design measurement studies that are representative of the real world as possible. 

\noindent\textbf{Implications for policymakers.}
We observe that private information confinement within the EU is low. 
GDPR has not stopped companies from collecting private information from the end users as long as their services are GDPR-compliant. 
In light of these findings, we hope that our study would encourage policymakers to further regulate how private information is used by and shared among the companies and how accountability can be truly guaranteed (\eg, which company should be held accountable if a device identifier was abused by targeted advertising while the majority of apps on that mobile device collect device identification information).

\noindent\textbf{Study limitations.}
Our study relies on the static and dynamic analysis, and layered security engines at the backend to identify and fingerprint that certain API calls lead to specific private information leakage. This prevents us to capture private information collection activities that happen at runtime but are not captured by the company's analytical infrastructure, which the on-device security engine relies on. Therefore, our work covers the lower bound of global private information collection activities.
Nevertheless, despite such limitations, our study provides the most comprehensive view of private data collection by Android apps to date and actionable insights.     

While this paper is based on measurements collected from a user base that is three orders of magnitude larger than previous work, our dataset is biased towards the end users of a single mobile security product, and therefore still presents some biases. 
For example, the distribution of devices used this study is not heavily screwed towards any specific region. However, the device distribution in Asia is skewed towards India and Japan and does not have as many devices in China which is one of the top countries/markets in terms of mobile users. 
In terms of the representativeness of the analyzed apps, it is challenging to ascertain the coverage of our study since it is infeasible to determine the total number of all Android apps, given such a fragmented ecosystem and many alternative markets.    
Still, by analyzing 2.1M apps this study is covering one of the largest sets of apps to date and is in line with the largest datasets collected by the academic community~\cite{allix2016androzoo}. 

Our analysis of \emph{data controllers} presented in Section~\ref{sec:data_controller} rely on the identification of the organizations behind PIC domains. As detailed in Section~\ref{sec:datasets} the mapping of domains to their owner organization relies on multiple data sources providing connections between the domains, the networks that host them as well as the organizations supposedly maintaining these resources. Such connections are cross-checked in the different data sources to compensate for inaccuracies in each of the sources. We also take a conservative approach and automatically discard all connections that are not seen in all data sources. This naturally hurts the number of domains to which we can map an organization. However, we favor the accuracy of the domain to owner organization mapping over its coverage. It is also important to note that some apps communicate with raw IP addresses instead of relying on domains. Moreover, we have seen that more than 97\% of these IP addresses refer to CDNs or hosting or cloud providers which hinder the identification of their owner organization.

Finally, while in this paper we studied how private information is collected from devices, and where this information flows to, our study does not allow us to understand how this information is acted upon by data controllers (\ie whether and how it is used to track users).
This remains an open question for the research community.

\section{Related Work}
\label{sec:related_work}

In this section, we selectively review previous studies on PHA characterization, Android permission system, private data leakages and prevention, and third party advertising and tracking services. We refer the readers to~\cite{tan2015securing,felt2011android,fang2014permission,nauman2010apex,xu2016toward,suarez2020eight} for in-depth studies and surveys on securing Android devices in general.

\noindent \textbf{PHA characterization.} Previous studies mainly focused on analyzing PHAs and systematically characterize them from various aspects such as evasion mechanism~\cite{faruki2014android}, installation methods~\cite{zhou2012dissecting}, malicious payloads~\cite{zhou2012dissecting}, repackaging mechanism~\cite{zhou2012detecting,suarez2018eight,lindorfer2014andradar}, behaviors~\cite{lindorfer2014andrubis,yang2014droidminer}, monetization~\cite{felt2011survey}, etc. These efforts shed light on how Android PHAs operate in the wild~\cite{zhou2012dissecting}, main incentives of mobile malware~\cite{felt2011survey,lindorfer2014andradar}, weaknesses of some of the popular mitigation solutions~\cite{faruki2014android}, etc. However, they did not discuss potential threats posed by information collection on mobile devices as these efforts center on app analysis and offer a less comprehensive view of the real device prevalence.

\noindent \textbf{Android permission system.} Android permission system has been extensively covered in the previous literature~\cite{barrera2010methodology,nauman2010apex,felt2011android,au2012pscout,backes2017artist}. These studies on Android permissions have  mainly leveraged static analysis techniques to understand the role of a given permission~\cite{au2012pscout,felt2011android,gao2015pmdroid}, potential privacy violation incurred by overprivileged apps~\cite{felt2011android,sarma2012android}, permission circumvention~\cite{reardon201950},  description-to-permission fidelity~\cite{qu2014autocog}, and improve mapping of Android permissions to framework/SDK API methods~\cite{backes2016demystifying,aafer2018precise}. Some recent research efforts also utilize dynamic analysis systems to distinguish and trace the permissions requested by apps at the runtime and those requested by the app’s core functionality~\cite{diamantaris2019reaper} and generate a more precise call graph enabling the system to extract the permission specification and improve the mapping~\cite{luo2020heap}. Our study complements these studies by showing the scales and the prevalence of private information collection in the real world devices.

\noindent \textbf{Third-party advertising and tracking services (ATSes).} There are two main approaches on studying third-party advertising and tracking services (ATSes). One approach leverages static tools to decompile apps and identified the embedded trackers from API calls and quantify various aspects of trackers~\cite{seneviratne2015measurement,binns2018third}. These methods offer a view of tracker behavior and prevalence from app perspective. Another approach is leveraging network traffic either captured on device or by ISP providers to provide insights into the mobile advertising and tracking ecosystem from an information flow  perspective~\cite{joe2015sponsoring,vallina2012breaking,razaghpanah2018apps,iordanou2018tracing}.

\noindent \textbf{PII leakage detection and protection.} The root cause of PII leakage is because the end users are presented with a set of required permissions by the apps but not how they handle the data after permissions are granted. Previous studies showed that mobile apps leak more privacy information than their web counterpart~\cite{papadopoulos2017long,leung2016should}. To this end, research efforts mainly focused on monitoring private information flows~\cite{enck2014taintdroid,sun2016taintart}, detecting potential privacy leaks by apps~\cite{gibler2012androidleaks,ren2018bug,reardon201950}, sensitive data leakage via third party libraries~\cite{stevens2012investigating,demetriou2016free,liu2019privacy}, privacy implication caused by targeted advertising in apps~\cite{book2015empirical}, private data leakage via network traffic analysis~\cite{ren2016recon,continella2017obfuscation}, impact of GDPR notices~\cite{utz2019informed}, and privacy implications incurred by pre-installed apps~\cite{gamba2020analysis}. Our work complements the previous work and shows a holistic picture of the state of sensitive information collection on Android in the wild, identifying the big players in this space (both legitimate companies and malicious actors), together with geographic trends.

\section{Conclusion}

In this paper, we presented the most comprehensive measurement study on private information collection on Android to date.
We showed that PIC is widespread on Android, and that various types of information are collected, with actors operating in different geographic areas interested in different types of information.
While most information flows terminate in the US, 7\% of the flows that we observe are directed to China.
We also find that data regulation laws like GDPR have not been effective in limiting the amount of personal information that flows to third countries \textbf{outside EU}.

\section*{Acknowledgments}

We wish to thank the anonymous reviewers for their feedback and our shepherd Adwait Nadkarni for his help in improving this paper.

\bibliographystyle{abbrv}
\bibliography{refs}

\end{document}